\documentclass[twocolumn]{aastex631}
\usepackage{amsmath}

\newcommand{\e}{\epsilon}

\newcommand{\g}{\gamma}

\newcommand{\psim}{\lower.5ex\hbox{$\; \buildrel \propto \over\sim \;$}}
\newcommand{\lbar}{\lower.0ex\hbox{$\; \buildrel{\lower0.0ex \hbox{-}} \over\lambda  \;$}}

\newcommand{\tgg}{\tau_{\gamma\gamma}}

\newcommand{\cm}{\mathrm{cm}}

\newcommand{\eV}{\mathrm{eV}}
\newcommand{\keV}{\mathrm{keV}}

\newcommand{\GeV}{\mathrm{GeV}}
\newcommand{\TeV}{\mathrm{TeV}}
\newcommand{\s}{\mathrm{s}}

\newcommand{\Kelvin}{\mathrm{K}}

\newcommand{\srad}{\mathrm{srad}}

\shorttitle{Lorentz Invariance Violation with Absorption of Gamma-rays by Solar Photons}
\shortauthors{Finke \& Patel}

\begin{document}

\title{Probing Lorentz Invariance Violation with Absorption of Astrophysical Gamma-rays by Solar Photons}

\author{Justin D.\ Finke}
\email{justin.finke@nrl.navy.mil}
\author{Parshad Patel}
\email{parshadkp@gmail.com}
\affiliation{%
U.S.\ Naval Research Laboratory, Code 7653, 4555 Overlook Ave.\ SW, Washington, DC 20375-5352, USA
}%

\date{\today}

\begin{abstract}
We compute in detail the absorption optical depth for astrophysical $\g$-ray photons interacting with solar photons to produce electron positron pairs.  This effect is greatest for $\g$-ray sources at small angular distances from the Sun, reaching optical depths as high as $\tau_{\g\g}\sim 10^{-2}$.  We also calculate this effect including modifications to the absorption cross section threshold from subluminal Lorentz invariance violation (LIV).  We show for the first time that subluminal LIV can lead to increases or decreases in $\tau_{\g\g}$ compared to the non-LIV case.  We show that, at least in principle,  LIV can be probed with this effect with observations of $\g$-ray sources near the Sun at $\gtrsim20\ \TeV$ by HAWC or LHAASO, although a measurement will be extremely difficult due to the small size of the effect.
\end{abstract}


\section{\label{sec:intro}Introduction}

Lorentz invariance is one of the fundamental principles of special relativity.  However, the violation of Lorentz invariance has been explored in various theories beyond the Standard Model, such as string theory, brane worlds, and loop quantum gravity \citep[e.g.,][]{amelino98,kifune99,amelino01,stecker01,mattingly05,christiansen06,jacobson06,ellis08}.  In the presence of Lorentz invariance violation (LIV), the normal relativistic dispersion relation for photons is modified as 
\begin{flalign}
E^2 - p^2c^2 = \pm E^2 \left(\frac{E}{ E_{\rm LIV}}\right)^n.
\end{flalign}
Here $E_{\rm LIV}$ is the energy scale, $n$ is an
integer (the order of the leading correction), and ``$+$'' represents
superluminal LIV, and ``$-$'' represents subluminal LIV.  For LIV brought about by quantum gravity models, it would be natural for $E_{\rm LIV}$ to be near the Planck Energy, $E_{\rm Planck} = 1.2 \times 10^{28}\ \eV$.  
LIV can lead to a number of potentially observable effects (for a review see \citep{mattingly05}), including a number of astrophysically interesting effects \citep[e.g.,][]{sarkar02,martinez20,desai23}.  One is that the speed of light is no longer constant, and is energy-dependent.  Using this effect, time-of-flight experiments can be used with astrophysical $\g$-ray transients (such as gamma-ray bursts [GRBs] and blazars) to constrain LIV \citep[e.g.,][]{abdo09_090510,vasil13,ellis19}.  The lack of photon decay in Galactic $\g$-ray sources measured with HAWC has led to strong constraints on superluminal LIV \citep{albert20}.  

Another effect is the modification of the threshold for pair production from photon-photon interactions, i.e., the process $\g + \g \rightarrow e^+ + e^-$.  Subluminal LIV can be constrained with this process with $\g$-rays from extragalactic sources (such as blazars and GRBs), which are absorbed by ultraviolet through far-infrared photons from the extragalactic background light \citep[EBL; e.g.,][]{nikishov62,gould67_EBL,fazio70,franceschini08, razzaque09, finke10_EBL, kneiske10, dominguez11, helgason12, stecker12, scully14, khaire15, stecker16, franceschini17, andrews18, khaire19, saldana21, finke22}.  The EBL is the integrated background light from all the stars that have existed in the observable universe, either through direct emission, or through absorption and re-radiation by dust.  Both the EBL and the $\g$-ray spectra of extragalactic sources are rather uncertain.  Nevertheless, after the discovery of photons out to 20 TeV from Mrk 501 by HEGRA \citep{aharonian99}, a number of authors suggested LIV was needed to explain how the $\g$-rays at these high energies could avoid EBL absorption and reach Earth \citep[e.g.,][]{kifune99,protheroe00}.  
Since then, studies of the $\g + \g \rightarrow e^+ + e^-$ process between extragalactic $\g$-ray sources and the EBL have been used to constrain LIV \citep[e.g.,][]{biteau15,tavecchio16,abdalla19,dzhappuev22,baktash22,li23,finke23}.  This will continue with the greatly improved sensitivity of the Cherenkov Telescope Array \citep[CTA;][]{abdalla21}.  Constraints have also been made from the upper limits on ultra-high energy $\g$-ray flux measured by the Auger Observatory \citep[e.g.,][]{lang18}.  In this manuscript, we explore only the effect of subluminal LIV on $\g\g$ absorption.  

There are (at least) two formulations of the effects of LIV on the photon-photon pair production process commonly used in the literature.  One is by \citet{jacob08}.  In their formulation, the standard model cross section formula is used, modifying only the threshold energy and the cross-section's dependence on this threshold.  The other is by \citet{fairbairn14}.  Those authors define an effective mass that is related to the LIV energy scale, and use that to define an invariant center of mass energy.   The two formulations {\em do not} give equivalent results.  A comparison of the two is given by \citet{tavecchio16}.  

Another possible application of the $\g + \g \rightarrow e^+ + e^-$ process to LIV constraints could come from the extinction of astrophysical $\g$-ray photons by photons from the Sun.  Preliminary estimates of this process were done by \citep{loeb22,balaji23}.  Here we explore this effect in much greater detail, including its application to constraining LIV.    As we will see below, subluminal LIV can make the $\g$-ray absorption optical depth more transparent {\em or more opaque} than it would be without LIV, depending on the photon distribution interacting with the $\g$-rays.  This may allow LIV to be constrained with measurements of the $\g$-ray sky at energies of $\gtrsim 20$\ TeV, although such measurements will be difficult to make.

In Section \ref{sec:formalism}, we describe our detailed calculations of absorption of astrophysical $\g$-rays by solar photons, both with and without LIV included.  In Section \ref{sec:LIV} we describe the potential for observing this effect and how it could be used to constrain LIV.  Finally we conclude with a summary in Section \ref{sec:discussion}.

\section{Formalism}
\label{sec:formalism}

\subsection{Setup}

\begin{figure}
\plotone{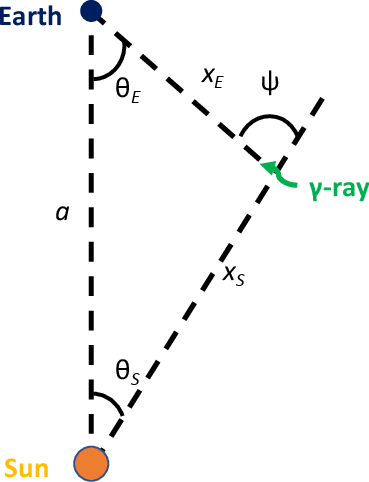}
\caption{\label{fig:geometry} Geometry of solar system and $\g$-ray heading towards Earth. Distances and angles are labeled.}
\end{figure}

Consider an astrophysical $\g$-ray photon coming towards the Earth from outside the solar system.  Let the angle at the Earth between the sun and the $\g$-ray photon be $\theta_E \equiv \cos^{-1}(\mu_E)$; the angle at the Sun between the Earth and the $\g$-ray photon is $\theta_S \equiv \cos^{-1}(\mu_S)$.  The distance between the Earth and the Sun is $a$; the distance between the Sun and the incoming $\g$-ray photon is $x_S$; and the distance between the Earth and the $\g$-ray photon is $x_E$.  See Fig.\ \ref{fig:geometry} for an illustration of the geometric setup.  

The $\g$-ray photon will interact with solar photons, annihilating the
photons and producing electron-positron pairs (i.e., $\g+ \g \rightarrow
e^+ + e^-$).  The $\g\g$ absorption cross section is \citep[e.g.,][]{gould67,brown73_PRD}
\begin{flalign}
\sigma_{\g\g}(s) & = \frac{1}{2}\pi r_e^2(1-\beta_{\rm cm}^2)
\Biggr[ (3-\beta_{\rm cm}^4)\ln\left(\frac{1+\beta_{\rm cm}}{1-\beta_{\rm cm}}\right)
\nonumber \\ &
-2\beta_{\rm cm}(2-\beta_{\rm cm}^2)\Biggr]
\end{flalign}
where
\begin{flalign}
\beta_{\rm cm} = \sqrt{1-s^{-1}}\ ;
\end{flalign}
$\sqrt{s}$ is the Lorentz factor of the resulting electron and positron in their center-of-momentum frame; and $r_e=2.817 \times 10^{-13}$ cm is the classical electron radius.

The $\g$-ray absorption optical depth per unit distance is given by \citep[e.g.,][]{dermer09_book}
\begin{flalign}
\label{taugg}
\frac{d\tau_{\g\g}(\e_1)}{dx_E} & = \int^{2\pi}_0d\phi \int^{1}_{-1}d\mu
(1-\cos\psi)
\nonumber \\ & \times
\int^{\infty}_0d\e\ n(\e,\Omega) 
\sigma_{\g\g}\left[\frac{\e\e_1(1-\cos\psi)}{2}\right]\ ,
\end{flalign}
where $\e_1$ is the dimensionless energy of the $\g$-rays; $\e$ is the dimensionless energy of the soft photons interacting with and absorbing the $\g$-rays; and $n(\e,\Omega)$ is the photon density of the photon field that interacts with and absorbs the $\g$-rays.    We use the notation where $\e_1$ and $\e$ represent dimensionless photon energies, i.e. photon energies in units of the electron rest energy, $m_e c^2=511\ \keV$.  Here $\psi$ is the angle between the incoming $\g$-ray photon and the photon from the Sun.  In general, the radiation field $n(\e,\Omega)$ and the interaction angle $\psi$ can be functions of both the polar angle ($\theta$) and the azimuthal angle ($\phi$).  However, for our geometry here they are independent of $\phi$. Some trigonometric effort yields
\begin{flalign}
\cos\psi =  \mu_S\mu_E - \sqrt{1-\mu_S^2}\sqrt{1-\mu_E^2}\ .
\end{flalign}
Note that \citet{balaji23} neglect the $1-\cos\psi$ term in Equation (\ref{taugg}).  This is the primary difference in our calculations and theirs.

The $\g$-ray comes from a source with intrinsic (i.e., unabsorbed) $\g$-ray flux $F_{\rm int}(\e_1)$.  It will be observed by a detector on or orbiting the Earth with a $\g$-ray flux $F_{\rm obs}(\e_1) = \exp(-\tau_{\g\g}(\e_1,\mu_E))F_{\rm int}(\e_1)$.

\subsection{Monochromatic Approximation for Sun}
\label{sec:monochromatic}

We follow the formalism of \cite{boettcher95,dermer09_book,dermer09,finke16} to derive formulae for the photoabsorption of astrophysical $\g$-rays  by solar photons as a function of $\g$-ray photon energy ($\e_1$) and angular distance on the sky of the source from the Sun ($\theta_E$).  In spherical coordinates $(R_r,\theta_r,\phi_r)$ are the radial distance, polar angle, and azimuthal angle, respectively, for the generic radiating medium, and $\mu_r = \cos\theta_r$.  In our case the radiating medium will be the Sun.  

We approximate the Sun as a point source emitting monochromatic photons with dimensionless energy $\e_S=2.7\Theta$ and luminosity $L_S=3.846 \times 10^{33}$ ergs s$^{-1}$.  Here $\Theta=k_BT_S/(m_ec^2)$ where $k_B=1.380 \times 10^{-16}$ erg K$^{-1}$ is the Boltzmann Constant and $T_S=5780\ \Kelvin$ is the effective temperature of the Sun.  The center of the Sun is at the origin (see Fig.\ \ref{fig:geometry}).  In this case, the emissivity of the photons emitted by the Sun is
\begin{flalign}
\label{emissivity1}
\dot n(\e,\Omega) = \frac{L_S}{m_e c^2 \e_S}
\frac{\delta(\e-\e_S)\delta(R_r-R_S)}{4\pi R_S^2}\ ,
\end{flalign}
where $R_S = 6.98 \times 10^{10}\ \cm$ is the radius of the Sun.  The photon density at a
distance $x$ from the origin is \citep[e.g.,][]{boettcher95,finke16}
\begin{flalign}
\label{density1}
n(\e,\Omega) & = \frac{1}{4\pi c}\int_0^{2\pi}d\phi_r \int^{1}_{-1}d\mu_r
\int_0^\infty dR_r \left(\frac{R_r}{x}\right)^2 
 \nonumber \\ & \times
\dot n(\e,\Omega) \delta(\phi_r-\phi_S)\delta(\mu-\mu_S)\ .
\end{flalign}
The coordinate system is chosen so that the azimuthal angle of the sun is $\phi_S=0$.  Putting Equation (\ref{emissivity1}) in
Equation (\ref{density1}) gives 
\begin{flalign}
\label{density2}
n(\e,\Omega) = \frac{L_S}{m_e c^2\e_Sc}\frac{1}{4\pi x^2}
\frac{\delta(\mu-\mu_S)}{2\pi} \delta(\e-\e_S)\ .
\end{flalign}
We are interested in the case where the $\g$-ray photon interacts with
a solar photon, which occurs at $x=x_S$.  Substituting Equation
(\ref{density2}) into Equation (\ref{taugg}) results in 
\begin{flalign}
\label{taugg1}
\tau_{\g\g}(\e_1,\mu_E) & = \frac{L_S}{4\pi\e_S m_e c^3}\int_0^\infty dx_E 
\frac{1-\cos\psi}{x_S^2} 
\nonumber \\  & \times
\sigma_{\g\g}\left[\frac{\e_S\e_1(1-\cos\psi)}{2}\right]\ ,
\end{flalign}
where, from the law of cosines, 
\begin{flalign}
x_S^2 = a^2+x_E^2-2ax_E\mu_E \ 
\end{flalign}
and
\begin{flalign}
\mu_S = \frac{a^2 + x_S^2 - x_E^2}{2ax_S}\ .
\end{flalign}
If $\theta_E = \pi$ then $\mu_E=-1$, $\mu_S=1$, and the integral in Equation
(\ref{taugg1}) can be performed analytically, giving 
\begin{flalign}
\tau_{\g\g}(\e_1,\mu_E=-1) = \frac{L_S}{2\pi \e_S m_e c^3a}
\sigma_{\g\g}(\e_S\e_1)\ .
\end{flalign}
This is a factor of 2 larger than the preliminary calculation by \citet{loeb22}.

\subsection{Blackbody Approximation for Sun}

Now we make the more realistic assumption that, instead of emitting
monochromatically, the sun emits as a blackbody.  We still assume the
Sun is a point source.  In this case, the emissivity of photons emitted by the Sun
is
\begin{flalign}
\dot n(\e,\Omega) = \frac{L_S}{m_e c^2\e}
\frac{\delta(R_r-R_S)}{4\pi R_S^2} \frac{15}{\Theta^4\pi^4}
\frac{\e^3}{\exp(\e/\Theta)-1}\ .
\end{flalign}

Following the same procedure as in Section \ref{sec:monochromatic}, we get 
\begin{flalign}
\label{taugg2}
\tau_{\g\g}(\e_1,\mu_E) & = \frac{L_S}{4m_e c^3} \frac{15}{\Theta^4 \pi^5}
\int_0^\infty d\e \frac{\e^2}{\exp(\e/\Theta)-1}
\nonumber \\  & \times
\int_0^\infty dx_E 
\frac{1-\cos\psi}{x_S^2} 
\nonumber \\  & \times
\sigma_{\g\g}\left[\frac{\e\e_1(1-\cos\psi)}{2}\right]\ .
\end{flalign}
Again, for $\theta_E=\pi$, the integral over $x_E$ is analytic, so that 
\begin{flalign}
\tau_{\g\g}(\e_1,\mu_E=-1) & = \frac{L_S}{2 m_e c^3a} \frac{15}{\Theta^4 \pi^5}
\nonumber \\  & \times
\int_0^\infty d\e \frac{\e^2}{\exp(\e/\Theta)-1}
\sigma_{\g\g}(\e\e_1)\ .
\end{flalign}

\begin{figure}
\plotone{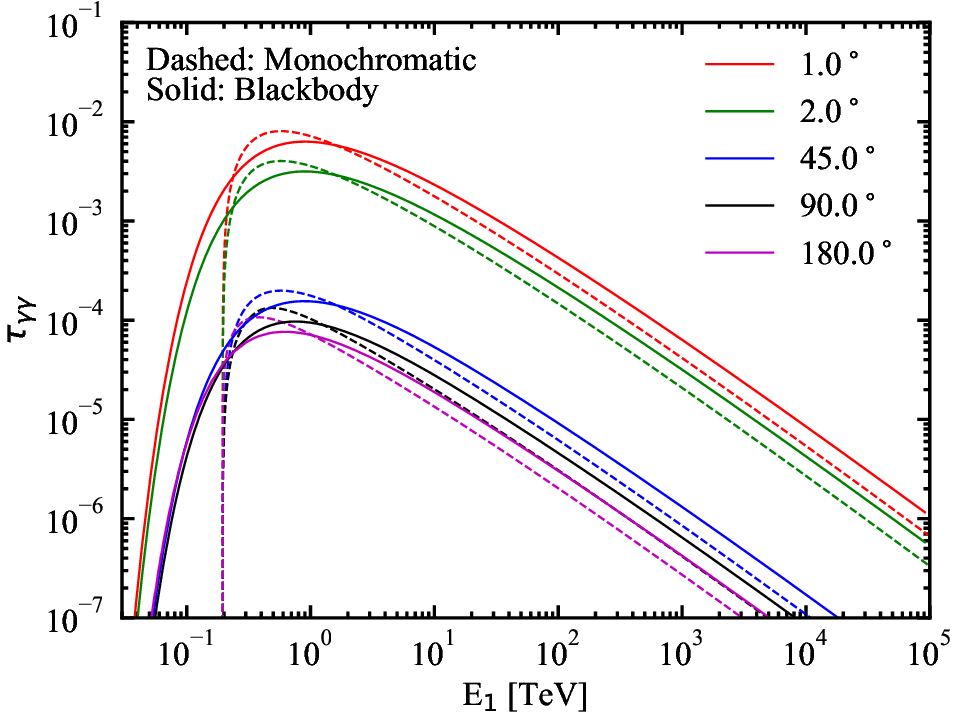}
\caption{\label{fig:MvsBB} Absorption optical depth for astrophysical $\g$-rays interacting with solar photons as a function of $\g$-ray energy, $E_1$, at different angular distances from the Sun ($\theta_E$) as indicated by the legend.  We show the monochromatic approximation (Equation [\ref{taugg1}]; dashed curves) and blackbody approximation (Equation [\ref{taugg2}]; solid curves).}
\end{figure}

In Figure \ref{fig:MvsBB} we plot $\tau_{\g\g}$ versus $\g$-ray energy $E_1=m_e c^2 \e_1$ for different angles from the Sun ($\theta_E$) for the monochromatic approximation (Equation [\ref{taugg1}]) and the blackbody approximation (Equation [\ref{taugg2}]).  As one can see, the results for the monochromatic and blackbody approximations are fairly similar, although the monochromatic approximation has a hard cutoff at about 200 GeV.  One cannot observe $\g$-ray photons that come from behind the Sun.  Since the Sun has an angular diameter of $\approx$0.5\textdegree{}, and solar-blind $\g$-ray telescopes typically have angular resolutions of $\sim1$\textdegree{}, one could not reasonably expect to detect an astrophysical $\g$-ray source at an angular distance of $\theta_E\lesssim$1\textdegree{}\ from the center of the Sun.

\subsection{Lorentz Invariance Violation}

Subluminal LIV can modify the threshold for the $\g\g$ absorption cross section, leading to an increase or decrease in $\tau_{\g\g}$ compared to the case without LIV.  There are (at least) two ways of implementing this used in the literature, that of \citet{jacob08} and \citet{fairbairn14}.  The two formulations do not give identical results.  

\subsubsection{Jacob \& Piran Formulation}

Following \cite{jacob08,biteau15}, in the presence of subluminal LIV, Equation (\ref{taugg1}) for the monochromatic approximation becomes 
\begin{flalign}
\tau_{\g\g}(\e_1,\mu_E) & = \frac{L_S}{4\pi \e_S m_e c^3}\int_0^\infty dx_E 
\frac{1-\cos\psi}{x_S^2} 
\nonumber \\ & \times
\sigma_{\g\g}\left[\frac{\e_S\e_1(1-\cos\psi)}
{2(1+0.25(\e_1/\e_{\rm LIV})^n\e_1^2)}\right]\ ,
\end{flalign}
where $\e_{\rm LIV}=E_{\rm LIV}/(m_e c^2)$.  Similarly for the blackbody approximation, Equation (\ref{taugg2}) becomes
\begin{flalign}
\tau_{\g\g}(\e_1,\mu_E) & = \frac{L_S}{4m_e c^3} \frac{15}{\Theta^4 \pi^5}
\int_0^\infty d\e \frac{\e^2}{\exp(\e/\Theta)-1}
\nonumber \\ & \times
\int_0^\infty dx_E 
\frac{1-\cos\psi}{x_S^2} 
\nonumber \\ & \times
\sigma_{\g\g}\left[\frac{\e\e_1(1-\cos\psi)}{2(1+0.25(\e_1/\e_{\rm LIV})^n\e_1^2)}\right]\ .
\label{BbLIV}
\end{flalign}

\begin{figure}
\plotone{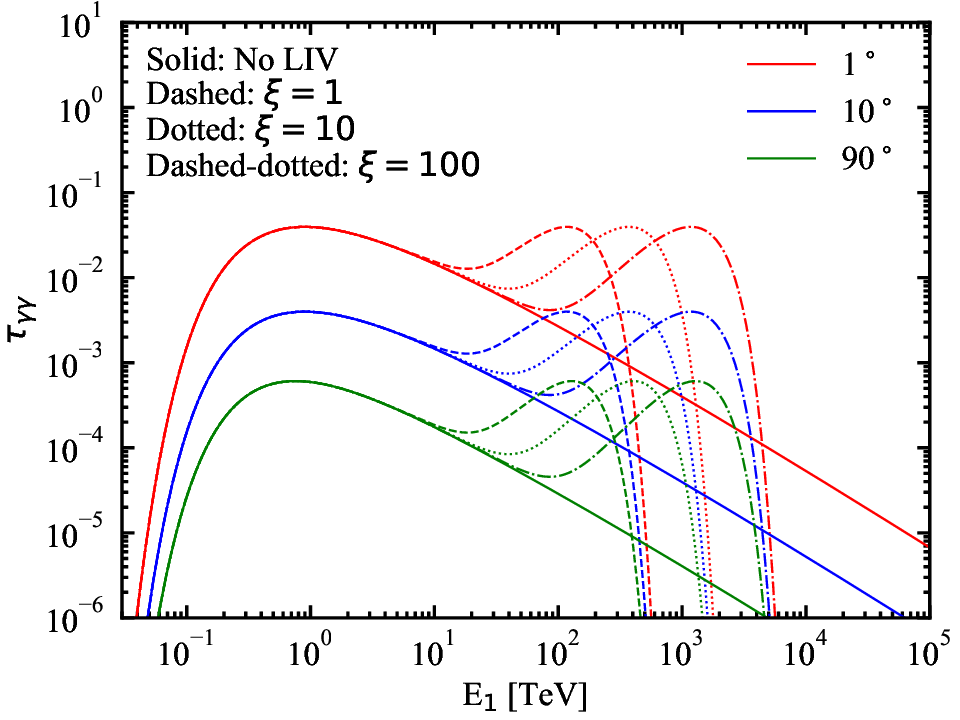}
\caption{\label{fig:BbLIV} Absorption optical depth for astrophysical $\g$-rays interacting with solar photons as a function of $\g$-ray energy, $E_1$, at different angular distances from the Sun ($\theta_E$) as indicated by the legend, using the blackbody approximation. We show the result without LIV (solid curves; Equation [\ref{taugg2}]) and with $n=1$ LIV using the \citet{jacob08} formulation (Equation [\ref{BbLIV}]) for $\xi=1$ (dashed curves), $\xi=10$ (dotted curves), and $\xi=100$ (dashed-dotted curves).  }
\end{figure}

We parameterize LIV with the parameter $\xi \equiv E_{\rm LIV}/E_{\rm Planck}$. In Figure \ref{fig:BbLIV} we show $\tau_{\g\g}$ plotted versus $E_1$ for $\theta_E=1$\textdegree, $\theta_E=10$\textdegree\ and $\theta_E=90$\textdegree\ for various values of $\xi$ in the linear ($n=1$) LIV case.  As the plot demonstrates, there is an {\em increase} in the gamma-ray absorption optical depth above 10 TeV, depending on the value of $\xi$.  This is because of the LIV effect modifying the argument for $\sigma_{\g\g}$.  With this modification, when sampling the cross section at energies where $0.25(\e_1/\e_{\rm LIV})^n\e_1^2 \gg 1$, one gets a cross section equivalent to the cross section at a lower energy.  As described by \citet{jacob08}:  ``[A]t any energy above $E^*$ [there is] an optical depth identical to the optical depth at some energy below $E^*$ for which the pair-production threshold is the same.''  In the notation of \citet{jacob08}, $E^*$ is the energy above which LIV effects become important for $\g\g$ absorption.  Thus, if the optical depth without LIV is {\em increasing} with energy, at energies where LIV is important, one will be sampling the optical depth at lower energies where $\tau_{\g\g}$ is lower, and thus one will find a {\em lower} $\tau_{\g\g}$ than without LIV.  However, if $\tau_{\g\g}$ is {\em decreasing} with energy, the LIV effect allows one to sample $\tau_{\g\g}$ at a lower energy where it is higher, and thus one will find a {\em higher} $\tau_{\g\g}$ than without LIV.

This is demonstrated in Fig.\ \ref{tau_EBL_test} for EBL absorption.  Here we plot the EBL absorption optical depth from the model of \citet{finke22} with no LIV, and with LIV for various values of $\xi$.  Here the LIV effect is computed as described by \citet{finke23}, following the formulation of \citet{jacob08}.  The dashed lines with arrows indicate the lower energies where $\tau_{\g\g}$ is being sampled for higher energies with LIV.  The plot demonstrates that when $\tau_{\g\g}$ is increasing, the LIV effect causes $\tau_{\g\g}$ to be lower than it otherwise would be.  But where $\tau_{\g\g}$ is decreasing with energy, as it is here above a few hundred TeV, the LIV can cause an increase in $\tau_{\g\g}$.

\begin{figure}
\plotone{tau_zeta_test_08}
\caption{\label{tau_EBL_test} EBL absorption optical depth versus $\g$-ray energy for the EBL model of \citet{finke22} at $z=0.15$.   This is shown without LIV (solid curve), and with $n=1$ LIV for the \citet{jacob08} formulation (dotted curves) and \citet{fairbairn14} formulation (dashed curves) for various values of $\xi$, as shown in the plot.  For the LIV curves from the \citet{jacob08} formulation, the dashed lines with arrows indicate the $\tau_{\g\g}$ at lower energies where the $\tau_{\g\g}$ at higher energies are sampling, due to the LIV effect.  Absorption by cosmic microwave background photons is not included in this plot.}
\end{figure}

\begin{figure}
\plotone{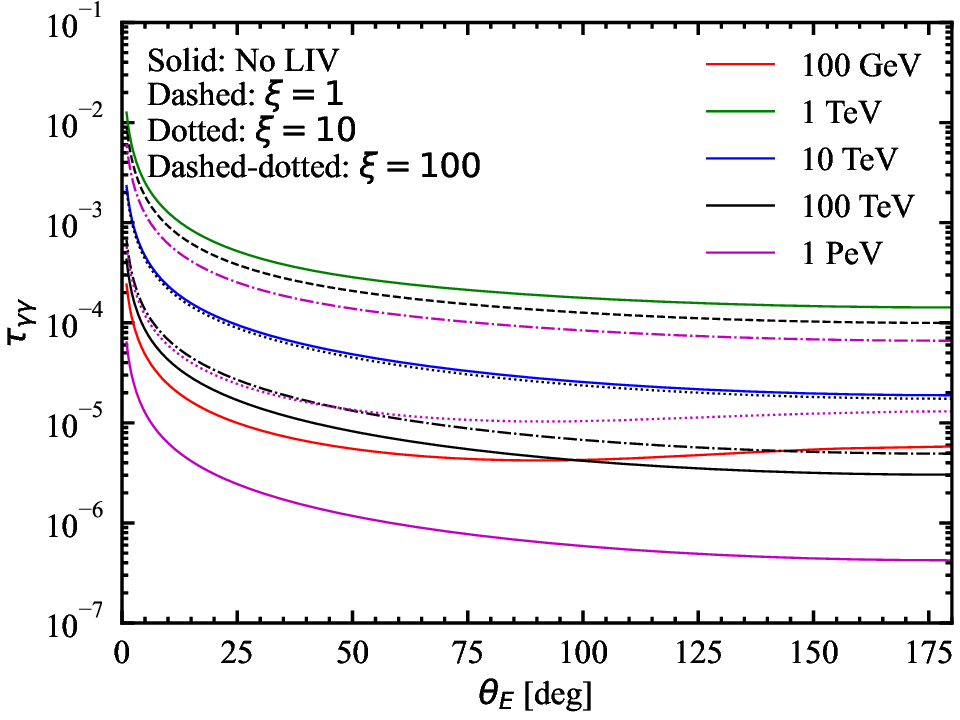}
\caption{\label{fig:BbLIVAngle} Absorption optical depth for astrophysical $\g$-rays interacting with solar photons as a function of angular distances from the Sun ($\theta_E$), for different $\g$-ray energies as indicated by the legend, using the blackbody approximation. We show the result without LIV (solid curves; Equation [\ref{taugg2}]) and with $n=1$ LIV using the \citet{jacob08} formulation (Equation [\ref{BbLIV}]) for $\xi=1$ (dashed curves), $\xi=10$ (dotted curves), and $\xi=100$ (dashed-dotted curves).  The 1 TeV curve has been scaled up by a factor of 2 and the 100 TeV $\xi=1$ curve has been scaled up by a factor of 1.5 for clarity.}
\end{figure}

\begin{figure}
\plotone{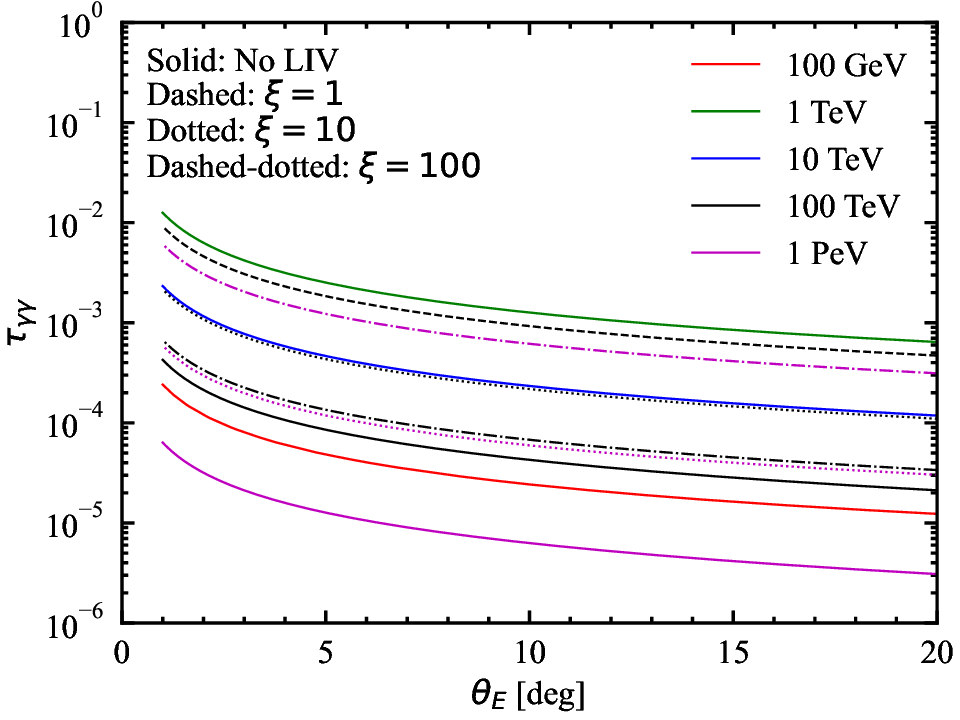}
\caption{\label{fig:BbLIVAngle20} The same as Fig.\ \ref{fig:BbLIVAngle}, but with $\theta_E$ from 0\textdegree\ to 20\textdegree.}
\end{figure}

In Figures \ref{fig:BbLIVAngle} and \ref{fig:BbLIVAngle20} we plot $\tgg$ versus $\theta_E$ for various $\g$-ray energies including effects from LIV.  The absorption optical depth decreases rapidly with increasing anglular distance from the Sun.  In most cases, the decrease in $\tau_{\g\g}$ is monotonic with angle.  This is because at large $x_E$ as $\theta_E$ increases, the cross section argument $\e_1\e(1-\cos\psi)/2$ increases, and the cross section generally decreases with increasing energy, as long as the invariant argument is above threshold.  However, the optical depth for 100 GeV decreases with increasing $\theta_E$ until it reaches a minimum at $\theta_E\approx75$\textdegree, then increases until $180$\textdegree.  In this case, the quantity $\e_1\e(1-\cos\psi)/2$ will go below threshold for parts of the integral over $x_E$ for smaller $\theta_E$. The exact results thus depend on the detailed kinematics.

For $E_1\lesssim10$\ TeV, there is no effect of LIV on $\g$-ray absorption, as shown in Figures \ref{fig:BbLIV} and \ref{fig:BbLIVAngle}.  For larger $E_1$, the absorption optical depth will ``mimic'' the absorption at lower energy when LIV is included, as described above.  Thus at 1 PeV with $\xi=10$, the result mimics the behavior of the 100 GeV curve, including the minimum at $\approx 75$\textdegree.  At 100 TeV for $\xi=1$ and 1 PeV for $\xi=10$ the results are identical to the 1 TeV curve.

\subsubsection{Fairbairn {\em et al.}\ Formulation}

\citet{fairbairn14} follow \citet{protheroe00} in their approach to LIV.  In their formulation, the absorption optical depth for astrophysical $\g$-rays due to interactions with solar photons is given by
\begin{flalign}
\tau_{\g\g}(\e_1,\mu_E) & = \frac{L_S}{4\pi \e_S m_e c^3}\int_0^\infty dx_E 
\frac{1-\cos\psi}{x_S^2} 
\nonumber \\ & \times
\sigma_{\g\g}\left[\frac{\e_S\e_1(1-\cos\psi)}{2}
- \frac{(\e_1/\e_{LIV})^n\e_1^2}{4}\right]\
\end{flalign}
with the monochromatic approximation, and
\begin{flalign}
\tau_{\g\g}(\e_1,\mu_E) & = \frac{L_S}{4m_e c^3} \frac{15}{\Theta^4 \pi^5}
\int_0^\infty d\e \frac{\e^2}{\exp(\e/\Theta)-1}
\nonumber \\ & \times
\int_0^\infty dx_E 
\frac{1-\cos\psi}{x_S^2} 
\nonumber \\ & \times
\sigma_{\g\g}\left[\frac{\e\e_1(1-\cos\psi)}{2}
- \frac{(\e_1/\e_{LIV})^n\e_1^2}{4}\right]\ 
\label{BbLIVFair}
\end{flalign}
with the blackbody approximation.

\begin{figure}
\plotone{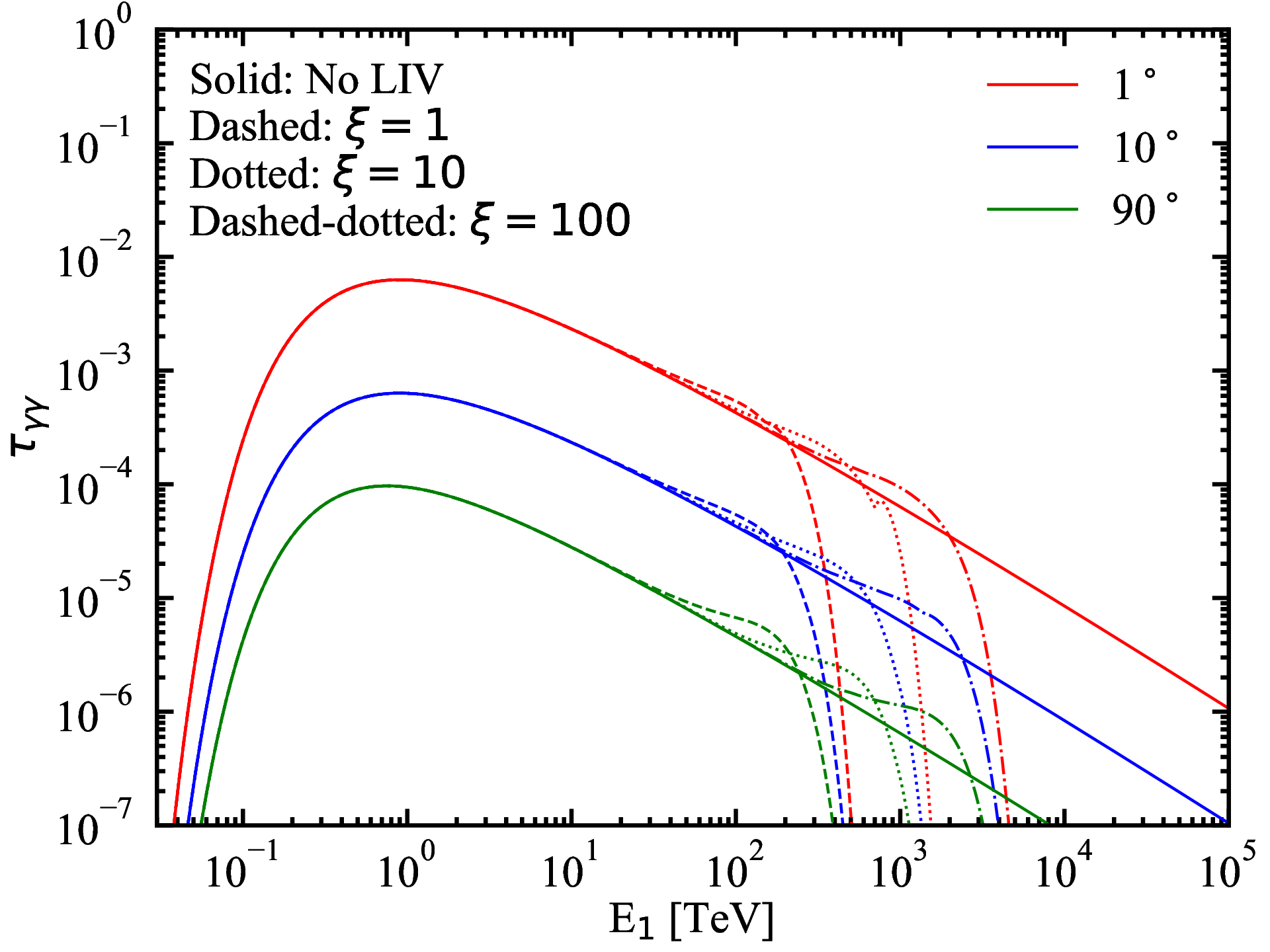}
\caption{\label{fig:BbLIVFair} Same as Figure \ref{fig:BbLIV}, but using the \citet{fairbairn14} formulation.}
\end{figure}

The absorption optical depth of astrophysical $\g$-rays with solar photons with LIV using the \citet{fairbairn14} formulation is shown in Fig.\ \ref{fig:BbLIVFair}; this can be compared with the \citet{jacob08} formulation plotted in Fig.\ \ref{fig:BbLIV}.  The Figure demonstrates that the calculation using the \citet{fairbairn14} formulation also can lead to an increase in the absorption optical depth relative to the calculation with no LIV, although the increase is much lower in the \citet{fairbairn14} formulation.  This is also true for absorption of $\g$-rays by EBL photons, as demonstrated in Fig.\ \ref{tau_EBL_test}; here again, on the region of the plot where $\tau_{\g\g}$ is decreasing with $E_1$ in the non-LIV case, the LIV can lead to an increase in $\tau_{\g\g}$ relative to the non-LIV case.  In general, the \citet{fairbairn14} formulation predicts lower absorption optical depths than the \citet{jacob08} formulation, as discussed by \citet{tavecchio16}.

\section{An Experiment for LIV}
\label{sec:LIV}

In the preceding section, we have computed the absorption optical depth of $\g$-ray photons from the Sun, both with and without LIV effects included.  These calculations make potentially observable predictions.  The absorption is generally greatest at small angular distances from the Sun.  This implies an experiment that could potentially detect or constrain LIV.  One would need to observe a bright $\g$-ray source when it is at small angular distances from the Sun, where the effect is strongest; and when it is quite far from the Sun, where the absorption is minimal.  The flux near the Sun would be 
\begin{flalign}
F({\rm near}) & = F({\rm far})\exp[-\tau_{\g\g}({\rm near})] 
\nonumber \\ &
\approx F({\rm far})[1-\tau_{\g\g}({\rm near})]\ ,
\end{flalign}
where $F({\rm far})$ is the flux far from the Sun.  The combination of these observations would give $\tau_{\g\g}({\rm near})$, the absorption optical depth near the Sun.  A comparison of this with our model predictions could then constrain $\xi$.  Assuming $F({\rm far})\approx F({\rm near}) \approx F$ for the purposes of estimating uncertainties, and $F({\rm far})$ and $F({\rm near})$ have the same uncertainty, $\sigma_F$, standard Gaussian error propagation gives the uncertainty in $\tau_{\g\g}({\rm near})$ as
\begin{flalign}
\label{errortau}
\sigma_{\tau_{\g\g}} = \sqrt{2}\left(\frac{\sigma_F}{F}\right)\ .
\end{flalign}

\begin{figure}
    \plotone{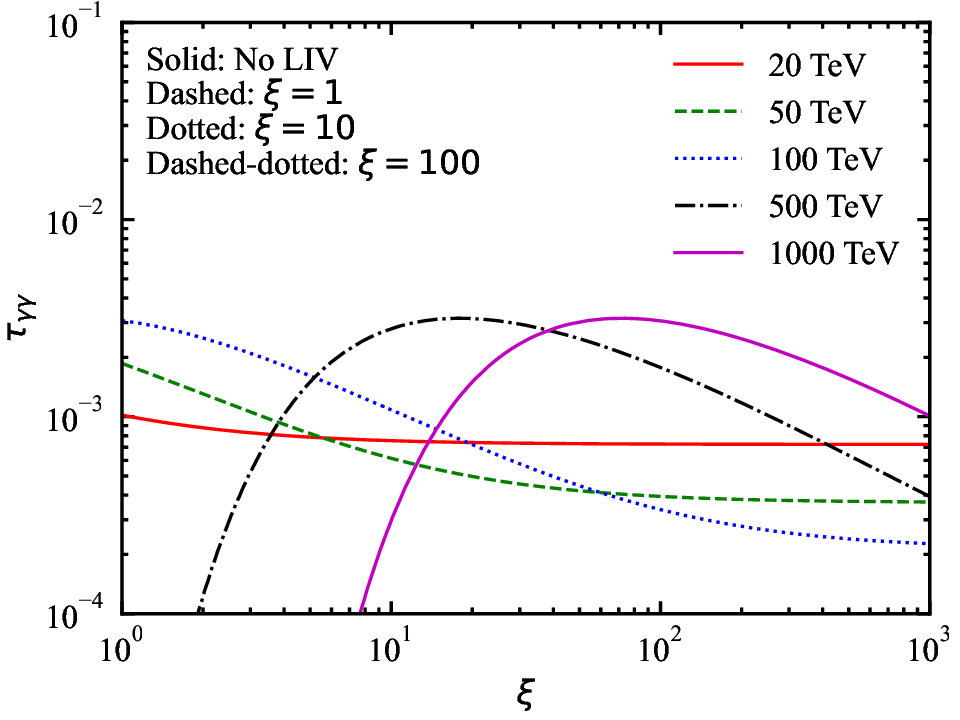}
    \caption{Absorption optical depth for astrophysical $\g$-rays interacting with solar photons as a function of $\xi$ for $n=1$ and $\theta_E=2$\textdegree, for different $\g$-ray energies, $E_1$, as indicated by the legend, using the blackbody approximation and the 
    \citet{jacob08} formulation of LIV.  }
    \label{fig:tauVsXi}
\end{figure}

In Fig. \ref{fig:tauVsXi} the absorption optical depth is plotted versus $\xi$ for $\theta_E=2$\textdegree\ for a variety of photon energies.  In general, observations of photons with higher energies have the greatest ability to constrain $\xi$.  Although note there is some ambiguity; a measurement of $\tau_{\g\g}=10^{-2}$ at 500 TeV could indicate $\xi\approx5$ or $\xi\approx 100$.  However, there are other ways to resolve these ambiguities; for instance, according to time-of-flight constraints from GRBs, $\xi\gtrsim10$ \citep{vasil13}.  To constrain LIV one must be able to observe a $\g$-ray source near the Sun at these very high energies.  Imaging Atmospheric Cherenkov Telescopes (IACTs) such as VERITAS, H.E.S.S., MAGIC, and the upcoming CTA are not able to observe during the day and thus could not observe $\g$-ray sources near the Sun, but could potentially observe this effect at large angles.  These leaves water Cherenkov detectors, such as HAWC and LHAASO.  {\em Fermi}-LAT could potentially measure the attenuation of $\g$-ray photons near the Sun, but it is not sensitive at energies relevant for constraining LIV.  

Compared with constraints on LIV from the EBL, the main advantage to this method is the higher degree of certainty.  With EBL absorption, neither the absorbing photon source (the EBL) nor the $\g$-ray source spectrum is completely known \citep[e.g.,][]{mattingly05}.  With the experiment outlined here, the absorbing photon source (the Sun) has a known spectrum that is approximated well by a blackbody.  The $\g$-ray source spectrum can be known from observations when the source is not near the Sun, when absorption by this effect is negligible.  The disadvantage of using the Sun is that the absorption is much smaller than with the EBL; for solar photons, $\tau_{\g\g}\sim 10^{-2}$ at most.  


We estimate the possibility of detecting absorption of astrophysical photons by the Sun at 1 TeV by HAWC and LHAASO, without regard to constraining LIV.  At $E_1=1$\ TeV and $\theta_E=2^\circ$, $\tau_{\g\g}\approx 3\times10^{-3}$.  Assuming Poisson uncertainties and a precisely known background,  $\sigma_F/F\approx \sqrt{S+B}/S$, where $S$ and $B$ are the source and background count rates, respectively, collected by the detector.  For HAWC and LHAASO we expect $B\gg S$, so that $\sqrt{S+B}/S\approx \sqrt{B}/S$.  Using Equation (\ref{errortau}), 
\begin{flalign}
\label{n1}
\frac{S}{\sqrt{B}} = \frac{\sqrt{2}}{\tau_{\g\g}} \left( \frac{\tau_{\g\g}}{\sigma_{\tau_{\g\g}}}\right)\ .
\end{flalign}
For a $3\sigma$ detection, $\tau_{\g\g}/\sigma_{\tau_{\g\g}}=3$.

One can search for the absorption of individual astrophysical $\g$-ray sources, or the $\g$-ray background.  The background has the advantage of always being observable near the Sun.  Its disadvantage is it is extremely faint. Indeed,
the extragalactic $\g$-ray background has not yet been detected directly (as far as the authors know) but estimates have been made by \cite{inoue16,qu19,qu22}.  

\startlongtable
\begin{deluxetable*}{cccccc}
\tablecaption{VHE Sources Within $10^\circ$ of the Sun. \label{table:sources}}
\tablewidth{0pt}
\tablehead{
\colhead{Name} &  \colhead{RA} &  \colhead{Dec} &  \colhead{Ecliptic Longitude [deg]} &  \colhead{Ecliptic Latitude [deg]} &  \colhead{Flux [Crab]} 
}
\startdata
\multicolumn{6}{c}{Sources from TevCat} \\
\hline
RGB J0152+017 & 1$^h$52$^m$33$^s$ & 1\textdegree46$'$40$''$ & 26.78 & -9.15 & 0.02 \\
1ES 0229+200 & 2$^h$32$^m$53$^s$ & 20\textdegree16$'$21$''$ & 42.29 & 4.99 & 0.018 \\
RBS 0413 & 3$^h$19$^m$47$^s$ & 18\textdegree45$'$42$''$ & 52.46 & 0.39 & 0.01 \\
VER J0521+211 & 5$^h$21$^m$45$^s$ & 21\textdegree12$'$51$''$ & 81.09 & -1.93 & 0.092 \\
Crab & 5$^h$34$^m$30$^s$ & 22\textdegree0$'$44$''$ & 84.09 & -1.30 & 1 \\
HAWC J0543+233 & 5$^h$43$^m$7.2$^s$ & 23\textdegree24$'$0.0$''$ & 86.13 & 0.02 & - \\
IC 443 & 6$^h$16$^m$51$^s$ & 22\textdegree30$'$11$''$ & 93.89 & -0.88 & 0.03 \\
Geminga & 6$^h$32$^m$28$^s$ & 17\textdegree22$'$0.0$''$ & 97.79 & -5.85 & 0.23 \\
RX J0648.7+1516 & 6$^h$48$^m$45$^s$ & 15\textdegree16$'$12$''$ & 101.86 & -7.67 & 0.033 \\
1ES 0647+250 & 6$^h$50$^m$46$^s$ & 25\textdegree3$'$0.0$''$ & 101.49 & 2.12 & 0.03 \\
2HWC J0700+143 & 7$^h$0$^m$28$^s$ & 14\textdegree19$'$12$''$ & 104.80 & -8.35 & - \\
2HWC J0819+157 & 8$^h$19$^m$55$^s$ & 15\textdegree47$'$24$''$ & 123.56 & -3.67 & - \\
RBS 0723 & 8$^h$47$^m$12$^s$ & 11\textdegree33$'$50$''$ & 131.06 & -6.12 & 0.025 \\
OJ 287 & 8$^h$54$^m$49$^s$ & 20\textdegree5$'$58$''$ & 130.51 & 2.59 & 0.013 \\
3C 279 & 12$^h$56$^m$11$^s$ & -5\textdegree47$'$22$''$ & 195.17 & 0.20 & - \\
2HWC J1309-054 & 13$^h$9$^m$14$^s$ & -5\textdegree29$'$24$''$ & 198.06 & 1.72 & - \\
PKS 1510-089 & 15$^h$12$^m$52$^s$ & -9\textdegree6$'$21$''$ & 228.30 & 8.49 & 0.03 \\
AP Librae & 15$^h$17$^m$41$^s$ & -24\textdegree22$'$19$''$ & 233.44 & -5.94 & 0.02 \\
TXS 1515-273 & 15$^h$18$^m$3.6$^s$ & -27\textdegree31$'$34$''$ & 234.35 & -8.96 & 0.06 \\
HESS J1741-302 & 17$^h$41$^m$15$^s$ & -30\textdegree22$'$37$''$ & 265.93 & -7.00 & 0.01 \\
HESS J1745-303 & 17$^h$45$^m$2.1$^s$ & -30\textdegree22$'$14$''$ & 266.75 & -6.97 & 0.05 \\
Galactic Centre & 17$^h$45$^m$39$^s$ & -29\textdegree0$'$22$''$ & 266.85 & -5.61 & 0.05 \\
HESS J1746-308 & 17$^h$46$^m$17$^s$ & -30\textdegree50$'$28$''$ & 267.03 & -7.44 & 0.03 \\
VER J1746-289 & 17$^h$46$^m$19$^s$ & -28\textdegree57$'$58$''$ & 267.00 & -5.56 & - \\
HESS J1746-285 & 17$^h$46$^m$23$^s$ & -28\textdegree52$'$33$''$ & 267.01 & -5.47 & - \\
MAGIC J1746.4-2853 & 17$^h$46$^m$25$^s$ & -28\textdegree52$'$55$''$ & 267.01 & -5.48 & - \\
SNR G000.9+00.1 & 17$^h$47$^m$23$^s$ & -28\textdegree9$'$6.0$''$ & 267.21 & -4.74 & 0.02 \\
Terzan 5 & 17$^h$47$^m$49$^s$ & -24\textdegree48$'$30$''$ & 267.23 & -1.40 & 0.015 \\
HESS J1800-240B & 18$^h$0$^m$26$^s$ & -24\textdegree2$'$20$''$ & 270.10 & -0.60 & - \\
W 28 & 18$^h$1$^m$42$^s$ & -23\textdegree20$'$6.0$''$ & 270.39 & 0.10 & - \\
HESS J1800-240A & 18$^h$1$^m$57$^s$ & -23\textdegree57$'$43$''$ & 270.45 & -0.52 & - \\
HESS J1804-216 & 18$^h$4$^m$31$^s$ & -21\textdegree42$'$0.0$''$ & 271.05 & 1.74 & 0.25 \\
HESS J1808-204 & 18$^h$8$^m$37$^s$ & -20\textdegree25$'$36$''$ & 272.02 & 3.00 & - \\
HESS J1809-193 & 18$^h$10$^m$31$^s$ & -19\textdegree18$'$0.0$''$ & 272.49 & 4.12 & 0.14 \\
HESS J1813-178 & 18$^h$13$^m$36$^s$ & -17\textdegree50$'$24$''$ & 273.25 & 5.56 & 0.06 \\
2HWC J1814-173 & 18$^h$14$^m$4.8$^s$ & -17\textdegree18$'$36$''$ & 273.38 & 6.09 & - \\
SNR G015.4+00.1 & 18$^h$18$^m$4.8$^s$ & -15\textdegree28$'$1.0$''$ & 274.40 & 7.90 & 0.018 \\
2HWC J1819-150* & 18$^h$19$^m$19$^s$ & -15\textdegree3$'$36$''$ & 274.71 & 8.30 & - \\
LHAASO J1825-1326 & 18$^h$25$^m$48$^s$ & -13\textdegree27$'$0.0$''$ & 276.37 & 9.85 & 3.57 \\
HESS J1825-137 & 18$^h$25$^m$49$^s$ & -13\textdegree46$'$35$''$ & 276.36 & 9.52 & 0.54 \\
2HWC J1825-134 & 18$^h$25$^m$50$^s$ & -13\textdegree24$'$0.0$''$ & 276.38 & 9.90 & - \\
LS 5039 & 18$^h$26$^m$15$^s$ & -14\textdegree49$'$30$''$ & 276.41 & 8.47 & 0.03 \\
1RXS J195815.6-301119 & 19$^h$58$^m$14$^s$ & -30\textdegree11$'$11$''$ & 295.61 & -9.34 & - \\
\hline
\tablebreak
\multicolumn{6}{c}{Additional Sources from LHAASO Catalog} \\
\hline
1LHAASO J0534+2200 & 5$^h$34$^m$26$^s$ & 22\textdegree3$'36''$ & 84.08 & -1.25 & - \\
1LHAASO J0542+2311 & 5$^h$42$^m$50$^s$ & 23\textdegree12$'0.0''$ & 86.06 & -0.18 & - \\
1LHAASO J0617+2234 & 6$^h$17$^m$24$^s$ & 22\textdegree35$'48''$ & 94.02 & -0.78 & - \\
1LHAASO J0634+1741 & 6$^h$34$^m$16$^s$ & 17\textdegree42$'36''$ & 98.20 & -5.49 & - \\
1LHAASO J0703+1405 & 7$^h$3$^m$19$^s$ & 14\textdegree6$'0.0''$ & 105.52 & -8.49 & - \\
1LHAASO J1809-1918 & 18$^h$9$^m$31$^s$ & -20\textdegree42'$0.0''$ & 272.23 & 2.72 & - \\
1LHAASO J1814-1719 & 18$^h$13$^m$4.0$^s$ & -18\textdegree7$'24''$ & 273.12 & 5.28 & - \\
1LHAASO J1814-1636 & 18$^h$14$^m$52$^s$ & -17\textdegree23$'12''$ & 273.57 & 6.01 & - \\
1LHAASO J1825-1418 & 18$^h$25$^m$9.0$^s$ & -15\textdegree41$'12''$ & 276.11 & 7.62 & - \\
1LHAASO J1825-1337 & 18$^h$25$^m$48$^s$ & -14\textdegree23$'48''$ & 276.32 & 8.90 & - 
\enddata
\end{deluxetable*}

The advantage of individual sources is they should have many more photons than the background.  The disadvantage is they do not stay near the Sun for long, since the Sun moves across the sky by approximately 1\textdegree\ day$^{-1}$.  This means a point source will only be observable with $10^\circ$ of the Sun for at most 20 days, so that for point sources it will be observed a fraction of the time $f_t=0.5 \times (20\ {\rm days})/(365\ {\rm days}) \approx 0.03$.  We searched the TeVCat (\url{http://tevcat.uchicago.edu}; \cite{wakely08}) for astrophysical TeV sources that get near the Sun.  In Table \ref{table:sources} we list the TeV sources with ecliptic latitude $<10$\textdegree.  As of this writing, the TeVCat has not included sources from the first LHAASO catalog \citep{cao23}.  We included these sources in our table as well.  The ecliptic latitude will be the sources' minimum distance on the sky from the Sun.  Variable sources such as blazars would not be ideal for these purposes, since one would want to compare their flux when they are near and far from the Sun.  Two sources stand out as the brightest:  the Crab \citep{aharonian21_lhaasocrab,cao21_lhaasocrab} and LHAASO J1825$-$1326 \citep{cao21}, seen by LHAASO up to $\approx 1$\ PeV and 500 TeV, respectively.  For point sources with photon flux $\Phi_S\equiv E\ dN/dE$ the signal count rate can be approximated as
\begin{flalign}
\label{Seqn}
S \approx \Phi_S A_{\rm eff} f_t \Delta t.
\end{flalign}
where $\Delta t$ is the length of time needed to observe the source and $A_{\rm eff}$ is the effective area of the detector.   For HAWC, we take the effective area from \citet{deyoung12}; for LHAASO we take it from \citet{cao19}.  Cosmic rays are a significant background for HAWC and LHAASO.  The analysis with HAWC is able to reject all but a fraction $f_r\approx6\times10^{-3}$ of these cosmic rays \citep{deyoung12}, and LHAASO can reject all but $f_r\approx2\times10^{-3}$ of them \citep{aharonian21_lhaasocrab}.  The cosmic ray background count rate is thus 
\begin{flalign}
\label{Beqn}
B \approx \Phi_{\rm CR} f_r A_{\rm eff} f_t \Delta t \Omega\ ,
\end{flalign}
where $\Phi_{\rm CR}$ is the $E\ dN/dE$ cosmic ray intensity and $\Omega$ is the solid angle around the source.  We take $\Phi_{\rm CR}$ from Chapter 29 of \citet{tanabashi18}.  For HAWC and LHAASO, with angular resolution of about $\theta_{\rm psf}\approx1\arcdeg$, $\Omega\approx \pi\theta_{\rm psf}^2 \approx 10^{-3}\ \srad$.  For a precisely known background, the number of counts above background is
\begin{flalign}
\label{n2}
\frac{S}{\sqrt{S+B}} \approx \frac{S}{\sqrt{B}} \approx \frac{\Phi_S}{\sqrt{\Phi_{\rm CR} f_r}}
\sqrt{ \frac{A_{\rm eff}f_t\Delta t}{\Omega}}\ .
\end{flalign}

\begin{deluxetable*}{lcc}
\tablecaption{Detection of $\tau_{\g\g}$ at 1 TeV by VHE $\gamma$-ray instruments. \label{table:VHEdetect}}
\tablehead{
\colhead{} &  \colhead{Crab} & \colhead{LHAASO J1825$-$1326}
 }
\startdata
$\Phi_S$ [$\cm^{-2}\ \s^{-1}$] & $4\times10^{-11}$ & $1.5\times10^{-11}$  \\
$\Phi_{\rm CR}$ [$\cm^{-2}\ \s^{-1}\ \srad^{-1}$] & $1.4\times10^{-5}$ & $1.4\times10^{-5}$  \\
$\tau_{\g\g}(\theta_E=2^\circ)$ & $3\times10^{-3}$ & $3\times10^{-3}$ \\
\hline
HAWC $f_r$ & $6\times10^{-3}$ & $6\times10^{-3}$ \\
HAWC $A_{\rm eff}$ [km$^2$] & 0.02 & 0.02  \\
HAWC $\Delta t$ [year] & $560$ & $4000$ \\
\hline
LHAASO $f_r$ & $2\times10^{-3}$ & $2\times10^{-3}$ \\
LHAASO $A_{\rm eff}$ [km$^2$] & 0.1 & 0.1  \\
LHAASO $\Delta t$ [year] & $37$ & $260$ \\
\enddata
\end{deluxetable*}

Combining Equations (\ref{n1}) and (\ref{n2}) gives the time needed for a detection,
\begin{flalign}
\Delta t \approx \frac{2}{\tau_{\g\g}^2}\left(\frac{\tau_{\g\g}}{\sigma_{\tau_{\g\g}}}\right)^2 \frac{\Phi_{\rm CR}f_r}{\Phi_S^2}\frac{\Omega}{A_{\rm eff} f_t}\ .
\end{flalign}
Here we estimate the time for detection for the Crab and LHAASO J1825$-$1326 at 1 TeV.  The results are in Table \ref{table:VHEdetect}.  In all cases, it would take $>30$ years to make a detection.  Note that although LHAASO J1825$-$1326 is listed as brighter than the Crab in Table \ref{table:sources}, the energy ranges here are different, so they are not directly comparable.  For the LHAASO J1825$-$1326 at 1 TeV listed in Table \ref{table:VHEdetect}, we have extrapolated the power-law spectrum from \citet{cao21} to lower energies.  

Next we look at whether {\em Fermi}-LAT could detect this effect.  This is complicated by the fact that for the LAT the Sun has a $\g$-ray halo around it extending for $\approx20$\textdegree\ caused by the Compton scattering of solar photons by cosmic ray electrons \citep{mosk06,orlando07,orlando08,abdo11_sun,linden22}.  The extragalactic gamma-ray background \citep[EGB;][]{ackermann15} becomes brighter than this halo at about 5\textdegree; we will assume the $\tau_{\g\g}$ at $\theta_E=10^\circ$, although the exact details will likely not effect our conclusions.  We use the LAT's Pass 8 on-axis instrument response function\footnote{\url{https://www.slac.stanford.edu/exp/glast/groups/canda/lat_Performance.htm}}.  The LAT is a sky survey instrument that sees the entire sky every 3 hours with a field of view of approximately 20\% of the sky.  So we take $f_t=0.2\times20/365=0.01$ for point sources.  The LAT has
an anti-coincidence detector to veto cosmic rays, so that at high energies it is signal dominated ($S\gg B$), and $\sigma_F/F \approx \sqrt{S+B}/S \approx \sqrt{S}/S = 1/\sqrt{S}$.  This leads to
\begin{flalign}
\label{dt_LAT}
\Delta t \approx \frac{2}{\Phi_S A_{\rm eff} f_t \tau_{\g\g}^2} 
\left(\frac{\tau_{\g\g}}{\sigma_{\tau_{\g\g}}}\right)^2\ .
\end{flalign}

\begin{deluxetable*}{lccc}
\tablecaption{Detection of $\tau_{\g\g}$ by {\em Fermi}-LAT 3FHL sources. \label{table:LATdetect}}
\tablehead{ & 
\colhead{50-150 GeV} & \colhead{150-500 GeV} & \colhead{500-2000 GeV} 
 }
\startdata
$\Phi_S$ & $3.9\times10^{-9}$ & $9.9\times10^{-10}$ & $2.3\times10^{-10}$ \\
$\tau_{\g\g}(\theta_E=10^\circ)$ & $2.4\times10^{-5}$ & $3.8\times10^{-4}$ & $6.3\times10^{-4}$ \\
$A_{\rm eff}$ [cm$^2$] & 9000 & 9000 & 9000 \\
$\Delta t$ [year] & $3\times10^9$ & $4\times10^7$ & $7\times10^7$ 
\enddata
\end{deluxetable*}

For individual sources, we searched the {\em Fermi}-LAT Third Hard Source Catalog \citep{ajell017_3fhl} for sources with ecliptic latitude $<10$\textdegree, i.e., sources that get within 10 degrees of the Sun, and found 253 sources.  Using Equation (\ref{dt_LAT}), we determine the total count rate for all of these sources for three energy bins given in the catalog:  the 50-150 GeV, 150-500 GeV, and 500-2000 GeV bins.  These results are in Table \ref{table:LATdetect}.  Even a stacking analysis of all hard LAT sources will not detect this effect in the likely lifetime of the {\em Fermi} spacecraft.  Further, most of these sources are blazars, which are often highly variable in $\g$-rays, and so would not be appropriate for this sort of measurement.

If HAWC or LHAASO could reject cosmic rays like the LAT, we could follow the same procedure for those experiments as we did for the LAT.  In that case we estimate HAWC and LHAASO would be able to detect this effect in the Crab at 1 TeV in 6.5 and 1.3 years, respectively.  Distinguishing between cosmic rays and photons is the strongest limitation on these experiments.

The prospects for detecting the absorption of $\g$-rays by solar photons at 1 TeV are poor.  Higher energies are more interesting from the point of view of constraining LIV, but prospects are even worse here, since fewer photons are detected.   We note that our approximation in this section is rather imprecise; we neglect all but Poisson uncertainties, and our approximations for source and background counts (Equations (\ref{Seqn}) and (\ref{Beqn})) do not properly integrate over energy.  However they are good enough for the order of magnitude level estimates provided here.

\section{Summary}
\label{sec:discussion}

In this paper:
\begin{itemize}
\item We compute in detail the effect of absorption of astrophysical $\g$-ray sources by solar photons, a process for which preliminary calculations were made by \citet{loeb22,balaji23}.  
\item We show the absorption effect is greatest for $\g$-ray sources at small angular distances from sun, reaching as high as $\tau_{\g\g}\sim10^{-2}$.
As pointed out by \citet{loeb22}, this effect causes the $\g$-ray background at $\gtrsim 100\ \GeV$ to have a greater anisotropy than the cosmic microwave background from the solar system's motion relative to the cosmic frame.  Unfortunately, the EGB is $\sim 10^6$ fainter than the CMB \citep[e.g.,][]{hauser01}, making the anisotropy much more difficult to detect.
\item We make calculations for this effect that include subluminal LIV.  For the first time, we show that subluminal LIV can lead to a decrease or increase in the absorption optical depth from the pair production process compared to the non-LIV case, depending on the spectrum of the absorbing photon source.  This is true for both the \citet{jacob08} and \citet{fairbairn14} formulation of LIV.  he \citet{jacob08} formulation predicts higher values of $\tau_{\g\g}$ than the \citet{fairbairn14} one.
\item We show that this effect is unlikely to be observed with water Cherenkov observatories like HAWC or LHAASO.  These can observe near the Sun, unlike IACTs, but the difficulties in separating $\g$-rays and hadrons means the needed precision will not be reached in a reasonable amount of time.   However, a clever stacking analysis may lead to a possible detection of the effect in a more reasonable amount of time.
\end{itemize}

\begin{acknowledgments}
We are grateful to the referee for comments that have improved this paper.  We would like to thank Teddy Cheung for pointing out there are sources in the LHAASO catalog that are not in the TeVCat at the time of this writing; and Brian Humensky for useful discussions on the observation of the effect described in this manuscript.  The authors were supported by NASA through contract S-15633Y.  J.D.F. was also supported by the Office of Naval Research.
\end{acknowledgments}





\bibliography{solar_absorb_ref, EBL_ref}{}

\begin{thebibliography}{}
\expandafter\ifx\csname natexlab\endcsname\relax\def\natexlab#1{#1}\fi

\bibitem[{{Abdalla} {et~al.}(2019){Abdalla}, {Aharonian}, {Ait Benkhali}, {Ang{\"u}ner}, {Arakawa}, {Arcaro}, {Armand}, {Arrieta}, {Backes}, {Barnard}, {Becherini}, {Becker Tjus}, {Berge}, {Bernhard}, {Bernl{\"o}hr}, {Blackwell}, {B{\"o}ttcher}, {Boisson}, {Bolmont}, {Bonnefoy}, {Bordas}, {Bregeon}, {Brun}, {Brun}, {Bryan}, {B{\"u}chele}, {Bulik}, {Bylund}, {Capasso}, {Caroff}, {Carosi}, {Cerruti}, {Chakraborty}, {Chandra}, {Chaves}, {Chen}, {Colafrancesco}, {Condon}, {Davids}, {Deil}, {Devin}, {deWilt}, {Dirson}, {Djannati-Ata{\"\i}}, {Dmytriiev}, {Donath}, {Doroshenko}, {O'C. Drury}, {Dyks}, {Egberts}, {Emery}, {Ernenwein}, {Eschbach}, {Fegan}, {Fiasson}, {Fontaine}, {Funk}, {F{\"u}{\ss}ling}, {Gabici}, {Gallant}, {Gat{\'e}}, {Giavitto}, {Glawion}, {Glicenstein}, {Gottschall}, {Grondin}, {Hahn}, {Haupt}, {Heinzelmann}, {Henri}, {Hermann}, {Hinton}, {Hofmann}, {Hoischen}, {Holch}, {Holler}, {Horns}, {Huber}, {Iwasaki}, {Jacholkowska}, {Jamrozy}, {Jankowsky}, {Jankowsky}, {Jouvin}, {Jung-Richardt},
  {Kastendieck}, {Katarzy{\'n}ski}, {Katsuragawa}, {Katz}, {Kerszberg}, {Khangulyan}, {Kh{\'e}lifi}, {King}, {Klepser}, {Klu{\'z}niak}, {Komin}, {Kosack}, {Krakau}, {Kraus}, {Kr{\"u}ger}, {Lamanna}, {Lau}, {Lefaucheur}, {Lemi{\`e}re}, {Lemoine-Goumard}, {Lenain}, {Leser}, {Lohse}, {Lorentz}, {L{\'o}pez-Coto}, {Lypova}, {Malyshev}, {Marandon}, {Marcowith}, {Mariaud}, {Mart{\'\i}-Devesa}, {Marx}, {Maurin}, {Meintjes}, {Mitchell}, {Moderski}, {Mohamed}, {Mohrmann}, {Moulin}, {Murach}, {Nakashima}, {de Naurois}, {Ndiyavala}, {Niederwanger}, {Niemiec}, {Oakes}, {O'Brien}, {Odaka}, {Ohm}, {Ostrowski}, {Oya}, {Padovani}, {Panter}, {Parsons}, {Perennes}, {Petrucci}, {Peyaud}, {Piel}, {Pita}, {Poireau}, {Priyana Noel}, {Prokhorov}, {Prokoph}, {P{\"u}hlhofer}, {Punch}, {Quirrenbach}, {Raab}, {Rauth}, {Reimer}, {Reimer}, {Renaud}, {Rieger}, {Rinchiuso}, {Romoli}, {Rowell}, {Rudak}, {Ruiz-Velasco}, {Sahakian}, {Saito}, {Sanchez}, {Santangelo}, {Sasaki}, {Schlickeiser}, {Sch{\"u}ssler}, {Schulz}, {Schwanke}, {Schwemmer},
  {Seglar-Arroyo}, {Senniappan}, {Seyffert}, {Shafi}, {Shilon}, {Shiningayamwe}, {Simoni}, {Sinha}, {Sol}, {Spanier}, {Specovius}, {Spir-Jacob}, {Stawarz}, {Steenkamp}, {Stegmann}, {Steppa}, {Takahashi}, {Tavernet}, {Tavernier}, {Taylor}, {Terrier}, {Tibaldo}, {Tiziani}, {Tluczykont}, {Trichard}, {Tsirou}, {Tsuji}, {Tuffs}, {Uchiyama}, {van der Walt}, {van Eldik}, {van Rensburg}, {van Soelen}, {Vasileiadis}, {Veh}, {Venter}, {Vincent}, {Vink}, {Voisin}, {V{\"o}lk}, {Vuillaume}, {Wadiasingh}, {Wagner}, {Wagner}, {White}, {Wierzcholska}, {Yang}, {Zaborov}, {Zacharias}, {Zanin}, {Zdziarski}, {Zech}, {Zefi}, {Ziegler}, {Zorn}, {{\.Z}ywucka}, \& {H.~E.~S.~S. Collaboration}}]{abdalla19}
{Abdalla}, H., {Aharonian}, F., {Ait Benkhali}, F., {et~al.} 2019, \apj, 870, 93

\bibitem[{{Abdalla} {et~al.}(2021){Abdalla}, {Abe}, {Acero}, {Acharyya}, {Adam}, {Agudo}, {Aguirre-Santaella}, {Alfaro}, {Alfaro}, {Alispach}, {Aloisio}, {Alves Batista}, {Amati}, {Amato}, {Ambrosi}, {Ang{\"u}ner}, {Araudo}, {Armstrong}, {Arqueros}, {Arrabito}, {Asano}, {Ascas{\'\i}bar}, {Ashley}, {Backes}, {Balazs}, {Balbo}, {Balmaverde}, {Baquero Larriva}, {Barbosa Martins}, {Barkov}, {Baroncelli}, {Barres de Almeida}, {Barrio}, {Batista}, {Becerra Gonz{\'a}lez}, {Becherini}, {Beck}, {Becker Tjus}, {Belmont}, {Benbow}, {Bernardini}, {Berti}, {Berton}, {Bertucci}, {Beshley}, {Bi}, {Biasuzzi}, {Biland}, {Bissaldi}, {Biteau}, {Blanch}, {Bocchino}, {Boisson}, {Bolmont}, {Bonanno}, {Bonneau Arbeletche}, {Bonnoli}, {Bordas}, {Bottacini}, {B{\"o}ttcher}, {Bozhilov}, {Bregeon}, {Brill}, {Brown}, {Bruno}, {Bruno}, {Bulgarelli}, {Burton}, {Buscemi}, {Caccianiga}, {Cameron}, {Capasso}, {Caprai}, {Caproni}, {Capuzzo-Dolcetta}, {Caraveo}, {Carosi}, {Carosi}, {Casanova}, {Cascone}, {Cauz}, {Cerny}, {Cerruti}, {Chadwick},
  {Chaty}, {Chen}, {Chernyakova}, {Chiaro}, {Chiavassa}, {Chytka}, {Conforti}, {Conte}, {Contreras}, {Coronado-Blazquez}, {Cortina}, {Costa}, {Costantini}, {Covino}, {Cristofari}, {Cuevas}, {D'Ammando}, {Daniel}, {Davies}, {Dazzi}, {De Angelis}, {de Bony de Lavergne}, {De Caprio}, {de C{\'a}ssia dos Anjos}, {de Gouveia Dal Pino}, {De Lotto}, {De Martino}, {de Naurois}, {de O{\~n}a Wilhelmi}, {De Palma}, {de Souza}, {Delgado}, {Della Ceca}, {della Volpe}, {Depaoli}, {Di Girolamo}, {Di Pierro}, {D{\'\i}az}, {D{\'\i}az-Bahamondes}, {Diebold}, {Djannati-Ata{\"\i}}, {Dmytriiev}, {Dom{\'\i}nguez}, {Donini}, {Dorner}, {Doro}, {Dournaux}, {Dwarkadas}, {Ebr}, {Eckner}, {Einecke}, {Ekoume}, {Els{\"a}sser}, {Emery}, {Evoli}, {Fairbairn}, {Falceta-Goncalves}, {Fegan}, {Feng}, {Ferrand}, {Fiandrini}, {Fiasson}, {Fioretti}, {Foffano}, {Fonseca}, {Font}, {Fontaine}, {Franco}, {Freixas Coromina}, {Fukami}, {Fukazawa}, {Fukui}, {Gaggero}, {Galanti}, {Gammaldi}, {Garcia}, {Garczarczyk}, {Gascon}, {Gaug}, {Gent}, {Ghalumyan},
  {Ghirlanda}, {Gianotti}, {Giarrusso}, {Giavitto}, {Giglietto}, {Giordano}, {Glicenstein}, {Goldoni}, {Gonz{\'a}lez}, {Gourgouliatos}, {Grabarczyk}, {Grandi}, {Granot}, {Grasso}, {Green}, {Grube}, {Gueta}, {Gunji}, {Halim}, {Harvey}, {Hassan Collado}, {Hayashi}, {Heller}, {Hern{\'a}ndez Cadena}, {Hervet}, {Hinton}, {Hiroshima}, {Hnatyk}, {Hnatyk}, {Hoffmann}, {Hofmann}, {Holder}, {Horan}, {H{\"o}randel}, {Horvath}, {Hovatta}, {Hrabovsky}, {Hrupec}, {Hughes}, {H{\"u}tten}, {Iarlori}, {Inada}, {Inoue}, {Insolia}, {Ionica}, {Iori}, {Jacquemont}, {Jamrozy}, {Janecek}, {Jim{\'e}nez Mart{\'\i}nez}, {Jin}, {Jung-Richardt}, {Jurysek}, {Kaaret}, {Karas}, {Karkar}, {Kawanaka}, {Kerszberg}, {Kh{\'e}lifi}, {Kissmann}, {Kn{\"o}dlseder}, {Kobayashi}, {Kohri}, {Komin}, {Kong}, {Kosack}, {Kubo}, {La Palombara}, {Lamanna}, {Lang}, {Lapington}, {Laporte}, {Lefaucheur}, {Lemoine-Goumard}, {Lenain}, {Leone}, {Leto}, {Leuschner}, {Lindfors}, {Lloyd}, {Lohse}, {Lombardi}, {Longo}, {Lopez}, {L{\'o}pez}, {L{\'o}pez-Coto},
  {Loporchio}, {Lucarelli}, {Luque-Escamilla}, {Lyard}, {Maggio}, {Majczyna}, {Makariev}, {Mallamaci}, {Mandat}, {Maneva}, {Manganaro}, {Manic{\`o}}, {Marcowith}, {Marculewicz}, {Markoff}, {Marquez}, {Mart{\'\i}}, {Martinez}, {Mart{\'\i}nez}, {Mart{\'\i}nez}, {Mart{\'\i}nez-Huerta}, {Maurin}, {Mazin}, {Mbarubucyeye}, {Medina Miranda}, {Meyer}, {Micanovic}, {Miener}, {Minev}, {Miranda}, {Mitchell}, {Mizuno}, {Mode}, {Moderski}, {Mohrmann}, {Molina}, {Montaruli}, {Moralejo}, {Morales Merino}, {Morcuende-Parrilla}, {Morselli}, {Mukherjee}, {Mundell}, {Murach}, {Muraishi}, {Nagai}, {Nakamori}, {Nemmen}, {Niemiec}, {Nieto}, {Nievas}, {Nikolajuk}, {Nishijima}, {Noda}, {Nosek}, {Nozaki}, {O'Brien}, {Ohira}, {Ohishi}, {Oka}, {Ong}, {Orienti}, {Orito}, {Orlandini}, {Orlando}, {Osborne}, {Ostrowski}, {Oya}, {Pagliaro}, {Palatka}, {Paneque}, {Pantaleo}, {Paredes}, {Parmiggiani}, {Patricelli}, {Pavleti{\'c}}, {Pe'er}, {Pech}, {Pecimotika}, {Peresano}, {Persic}, {Petruk}, {Pfrang}, {Piatteli}, {Pietropaolo}, {Pillera},
  {Pilszyk}, {Pimentel}, {Pintore}, {Pita}, {Pohl}, {Poireau}, {Polo}, {Prado}, {Prast}, {Principe}, {Produit}, {Prokoph}, {Prouza}, {Przybilski}, {Pueschel}, {P{\"u}hlhofer}, {Pumo}, {Punch}, {Queiroz}, {Quirrenbach}, {Rando}, {Razzaque}, {Rebert}, {Recchia}, {Reichherzer}, {Reimer}, {Reimer}, {Renier}, {Reposeur}, {Rhode}, {Ribeiro}, {Rib{\'o}}, {Richtler}, {Rico}, {Rieger}, {Rizi}, {Rodriguez}, {Rodriguez Fernandez}, {Rodriguez Ramirez}, {Rodr{\'\i}guez V{\'a}zquez}, {Romano}, {Romeo}, {Roncadelli}, {Rosado}, {Rosales de Leon}, {Rowell}, {Rudak}, {Rujopakarn}, {Russo}, {Sadeh}, {Saha}, {Saito}, {Salesa Greus}, {Sanchez}, {S{\'a}nchez-Conde}, {Sangiorgi}, {Sano}, {Santander}, {Santos}, {Sanuy}, {Sarkar}, {Saturni}, {Sawangwit}, {Scherer}, {Schleicher}, {Schovanek}, {Schussler}, {Schwanke}, {Sciacca}, {Scuderi}, {Seglar Arroyo}, {Sergijenko}, {Servillat}, {Seweryn}, {Shalchi}, {Sharma}, {Shellard}, {Siejkowski}, {Sinha}, {Sliusar}, {Slowikowska}, {Sokolenko}, {Sol}, {Specovius}, {Spencer}, {Spiga},
  {Stamerra}, {Stani{\v{c}}}, {Starling}, {Stolarczyk}, {Straumann}, {Stri{\v{s}}kovi{\'c}}, {Suda}, {{\'S}wierk}, {Tagliaferri}, {Takahashi}, {Takahashi}, {Tavecchio}, {Taylor}, {Tejedor}, {Temnikov}, {Terrier}, {Terzic}, {Testa}, {Tian}, {Tibaldo}, {Tonev}, {Torres}, {Torresi}, {Tosti}, {Tothill}, {Tovmassian}, {Travnicek}, {Truzzi}, {Tuossenel}, {Umana}, {Vacula}, {Vagelli}, {Valentino}, {Vallage}, {Vallania}, {van Eldik}, {Varner}, {Vassiliev}, {V{\'a}zquez Acosta}, {Vecchi}, {Veh}, {Vercellone}, {Vergani}, {Verguilov}, {Vettolani}, {Viana}, {Vigorito}, {Vitale}, {Vorobiov}, {Vovk}, {Vuillaume}, {Wagner}, {Walter}, {Watson}, {White}, {White}, {Wiemann}, {Wierzcholska}, {Will}, {Williams}, {Wischnewski}, {Wolter}, {Yamazaki}, {Yanagita}, {Yang}, {Yoshikoshi}, {Zacharias}, {Zaharijas}, {Zaric}, {Zavrtanik}, {Zavrtanik}, {Zdziarski}, {Zech}, {Zechlin}, {Zhdanov}, \& {{\v{Z}}ivec}}]{abdalla21}
{Abdalla}, H., {Abe}, H., {Acero}, F., {et~al.} 2021, Journal of Cosmology and Astro-Particle Physics, 2021, 048

\bibitem[{{Abdo} {et~al.}(2009){Abdo}, {Ackermann}, {Ajello}, {Asano}, {Atwood}, {Axelsson}, {Baldini}, {Ballet}, {B arbiellini}, {Baring}, {Bastieri}, {Bechtol}, {Bell azzini}, {Berenji}, {Bhat}, {Bissaldi}, {Bloom}, {Bonamente}, {Bonnell}, {Borgland}, {Bouvier}, {Bregeon}, {Brez}, {Briggs}, {Brigida}, {Bruel}, {Burgess}, {Burnett}, {Caliandro}, {Cameron }, {Caraveo}, {Casandjian}, {Cecchi}, {{\c{C} }elik}, {Chaplin}, {Charles}, {Cheung}, {Chiang }, {Ciprini}, {Claus}, {Cohen-Tanugi}, {Cominsky}, {Connaughton}, {Conrad}, {Cutini}, {Dermer}, {de Angelis}, {de Palma}, {Digel}, {Dingus}, {D o Couto E Silva}, {Drell}, {Dubois}, {Dumora}, {Far nier}, {Favuzzi}, {Fegan}, {Finke}, {Fishman}, {Foschini}, {Fukazawa}, {Funk}, {Fusco}, {Gargano}, {Gasparrini}, {Gehrels}, {Germani}, {Giebels}, {Giglietto}, {Giordano}, {Glanzm an}, {Godfrey}, {Granot}, {Greiner}, {Grenier}, {Grondin}, {Grove}, {Grupe}, {Guillemot}, { Guiriec}, {Hanabata}, {Harding}, {Hayashida}, {Hays }, {Hoversten}, {Hughes}, {J{\'o}hannesson}, {Jo
  hnson}, {Johnson}, {Johnson}, {Kamae}, {Katag iri}, {Kataoka}, {Kawai}, {Kerr}, {Kippen}, {Kn{\"o}dlseder}, {Kocevski}, {Kouveliotou}, {Kuehn}, {Kuss}, {Lande}, {Latronico}, {Lemoine-Goumard}, {Long o}, {Loparco}, {Lott}, {Lovellette}, {Lubrano}, {Madejski}, {Makeev}, {Mazziotta}, {McBreen}, {McGlynn}, {M{\'e}sz{\'a}ros}, {Meurer}, {Mitthumsiri}, {Mizuno}, {Moiseev}, {Monte}, {Monzani}, {Moretti}, {Morselli}, {Mos kalenko}, {Murgia}, {Nakamori}, {Nolan}, {Norris }, {Nuss}, {Ohno}, {Ohsugi}, {Omodei}, {Orla ndo}, {Ormes}, {Ozaki}, {Paciesas}, {Paneque}, {Panetta}, {Parent}, {Pelassa}, {Pepe}, {Pesc e-Rollins}, {Petrosian}, {Piron}, {Porter}, {Preece }, {Rain{\`o}}, {Ramirez-Ruiz}, {Rando}, {Razzano}, {Razzaque}, {Reimer}, {Reimer}, {Reposeur}, {Ritz }, {Rochester}, {Rodriguez}, {Roth}, {Ryde}, {Sadrozinski}, {Sanchez}, {Sander}, {Saz Parkinson }, {Scargle}, {Schalk}, {Sgr{\`o}}, {Siskind}, {Smith}, {Smith}, {Spandre}, {Spinelli}, {Stamatikos}, {Stecker}, {Strickman}, {Suson}, {Tajima}, {Takahashi},
  {Takahashi}, {Tanaka}, { Thayer}, {Thayer}, {Thompson}, {Tibaldo}, {To ma}, {Torres}, {Tosti}, {Troja}, {Uchiyama}, {Uehara}, {Usher}, {van der Horst}, {Vasileiou}, {Vitale}, {von Kienlin}, {Waite}, {Wan g}, {Wilson-Hodge}, {Winer}, {Wood}, {Wu}, {Yamazaki}, {Ylinen}, {Ziegler}, \& {Fermi LAT Collaboratio n}}]{abdo09_090510}
{Abdo}, A.~A., {Ackermann}, M., {Ajello}, M., {et~al.} 2009, \nat, 462, 331

\bibitem[{{Abdo} {et~al.}(2011){Abdo}, {Ackermann}, {Ajello}, {Baldini}, {Ballet}, {Barbiellini}, {Bastieri}, {Bechtol}, {Bellazzini}, {Berenji}, {Bonamente}, {Borgland}, {Bouvier}, {Bregeon}, {Brez}, {Brigida}, {Bruel}, {Buehler}, {Buson}, {Caliandro}, {Cameron}, {Caraveo}, {Casandjian}, {Cecchi}, {Charles}, {Chekhtman}, {Chiang}, {Ciprini}, {Claus}, {Cohen-Tanugi}, {Conrad}, {Cutini}, {de Angelis}, {de Palma}, {Dermer}, {Digel}, {Silva}, {Drell}, {Dubois}, {Favuzzi}, {Fegan}, {Focke}, {Fortin}, {Frailis}, {Funk}, {Fusco}, {Gargano}, {Gasparrini}, {Gehrels}, {Germani}, {Giglietto}, {Giordano}, {Giroletti}, {Glanzman}, {Godfrey}, {Grenier}, {Grillo}, {Guiriec}, {Hadasch}, {Hays}, {Hughes}, {Iafrate}, {J{\'o}hannesson}, {Johnson}, {Johnson}, {Kamae}, {Katagiri}, {Kataoka}, {Kn{\"o}dlseder}, {Kuss}, {Lande}, {Latronico}, {Lee}, {Lionetto}, {Longo}, {Loparco}, {Lott}, {Lovellette}, {Lubrano}, {Makeev}, {Mazziotta}, {McEnery}, {Mehault}, {Michelson}, {Mitthumsiri}, {Mizuno}, {Moiseev}, {Monte}, {Monzani},
  {Morselli}, {Moskalenko}, {Murgia}, {Nakamori}, {Naumann-Godo}, {Nolan}, {Norris}, {Nuss}, {Ohsugi}, {Okumura}, {Omodei}, {Orlando}, {Ormes}, {Ozaki}, {Paneque}, {Pelassa}, {Pesce-Rollins}, {Pierbattista}, {Piron}, {Porter}, {Rain{\`o}}, {Rando}, {Razzano}, {Reimer}, {Reimer}, {Reposeur}, {Ritz}, {Sadrozinski}, {Schalk}, {Sgr{\`o}}, {Share}, {Siskind}, {Smith}, {Spandre}, {Spinelli}, {Strickman}, {Strong}, {Takahashi}, {Tanaka}, {Thayer}, {Thayer}, {Thompson}, {Tibaldo}, {Torres}, {Tosti}, {Tramacere}, {Troja}, {Uchiyama}, {Usher}, {Vandenbroucke}, {Vasileiou}, {Vianello}, {Vilchez}, {Vitale}, {Vladimirov}, {Waite}, {Wang}, {Winer}, {Wood}, {Yang}, \& {Ziegler}}]{abdo11_sun}
---. 2011, \apj, 734, 116

\bibitem[{{Ackermann} {et~al.}(2015){Ackermann}, {Ajello}, {Albert}, {Atwood}, {Baldini}, {Ballet}, {Barbiellini}, {Bastieri}, {Bechtol}, {Bellazzini}, {Bissaldi}, {Blandford}, {Bloom}, {Bottacini}, {Brandt}, {Bregeon}, {Bruel}, {Buehler}, {Buson}, {Caliandro}, {Cameron}, {Caragiulo}, {Caraveo}, {Cavazzuti}, {Cecchi}, {Charles}, {Chekhtman}, {Chiang}, {Chiaro}, {Ciprini}, {Claus}, {Cohen-Tanugi}, {Conrad}, {Cuoco}, {Cutini}, {D'Ammando}, {de Angelis}, {de Palma}, {Dermer}, {Digel}, {Silva}, {Drell}, {Favuzzi}, {Ferrara}, {Focke}, {Franckowiak}, {Fukazawa}, {Funk}, {Fusco}, {Gargano}, {Gasparrini}, {Germani}, {Giglietto}, {Giommi}, {Giordano}, {Giroletti}, {Godfrey}, {Gomez-Vargas}, {Grenier}, {Guiriec}, {Gustafsson}, {Hadasch}, {Hayashi}, {Hays}, {Hewitt}, {Ippoliti}, {Jogler}, {J{\'o}hannesson}, {Johnson}, {Johnson}, {Kamae}, {Kataoka}, {Kn{\"o}dlseder}, {Kuss}, {Larsson}, {Latronico}, {Li}, {Li}, {Longo}, {Loparco}, {Lott}, {Lovellette}, {Lubrano}, {Madejski}, {Manfreda}, {Massaro}, {Mayer}, {Mazziotta},
  {McEnery}, {Michelson}, {Mitthumsiri}, {Mizuno}, {Moiseev}, {Monzani}, {Morselli}, {Moskalenko}, {Murgia}, {Nemmen}, {Nuss}, {Ohsugi}, {Omodei}, {Orlando}, {Ormes}, {Paneque}, {Panetta}, {Perkins}, {Pesce-Rollins}, {Piron}, {Pivato}, {Porter}, {Rain{\`o}}, {Rando}, {Razzano}, {Razzaque}, {Reimer}, {Reimer}, {Reposeur}, {Ritz}, {Romani}, {S{\'a}nchez-Conde}, {Schaal}, {Schulz}, {Sgr{\`o}}, {Siskind}, {Spandre}, {Spinelli}, {Strong}, {Suson}, {Takahashi}, {Thayer}, {Thayer}, {Tibaldo}, {Tinivella}, {Torres}, {Tosti}, {Troja}, {Uchiyama}, {Vianello}, {Werner}, {Winer}, {Wood}, {Wood}, {Zaharijas}, \& {Zimmer}}]{ackermann15}
{Ackermann}, M., {Ajello}, M., {Albert}, A., {et~al.} 2015, \apj, 799, 86

\bibitem[{{Aharonian} {et~al.}(2021){Aharonian}, {An}, {Axikegu}, {Bai}, {Bai}, {Bao}, {Bastieri}, {Bi}, {Bi}, {Cai}, {Cai}, {Cao}, {Cao}, {Chang}, {Chang}, {Chang}, {Chen}, {Chen}, {Chen}, {Chen}, {Chen}, {Chen}, {Chen}, {Chen}, {Chen}, {Chen}, {Chen}, {Chen}, {Chen}, {Cheng}, {Cheng}, {Cui}, {Cui}, {Cui}, {Dai}, {Dai}, {Dai}, {Danzengluobu}, {Della Volpe}, {Piazzoli}, {Dong}, {Fan}, {Fan}, {Fan}, {Fang}, {Fang}, {Feng}, {Feng}, {Feng}, {Feng}, {Gao}, {Gao}, {Gao}, {Gao}, {Ge}, {Geng}, {Gong}, {Gou}, {Gu}, {Guo}, {Guo}, {Guo}, {Guo}, {Han}, {He}, {He}, {He}, {He}, {He}, {He}, {Heller}, {Hor}, {Hou}, {Hou}, {Hu}, {Hu}, {Hu}, {Hu}, {Huang}, {Huang}, {Huang}, {Huang}, {Huang}, {Ji}, {Ji}, {Jia}, {Jiang}, {Jiang}, {Jin}, {Kuleshov}, {Levochkin}, {Li}, {Li}, {Li}, {Li}, {Li}, {Li}, {Li}, {Li}, {Li}, {Li}, {Li}, {Li}, {Li}, {Li}, {Li}, {Li}, {Li}, {Liang}, {Liang}, {Lin}, {Liu}, {Liu}, {Liu}, {Liu}, {Liu}, {Liu}, {Liu}, {Liu}, {Liu}, {Liu}, {Liu}, {Liu}, {Liu}, {Liu}, {Liu}, {Long}, {Lu}, {Lv}, {Ma}, {Ma}, {Ma},
  {Mao}, {Masood}, {Mitthumsiri}, {Montaruli}, {Nan}, {Pang}, {Pattarakijwanich}, {Pei}, {Qi}, {Qiao}, {Ruffolo}, {Rulev}, {S{\'a}iz}, {Shao}, {Shchegolev}, {Sheng}, {Shi}, {Song}, {Stenkin}, {Stepanov}, {Sun}, {Sun}, {Sun}, {Tam}, {Tang}, {Tian}, {Wang}, {Wang}, {Wang}, {Wang}, {Wang}, {Wang}, {Wang}, {Wang}, {Wang}, {Wang}, {Wang}, {Wang}, {Wang}, {Wang}, {Wang}, {Wang}, {Wang}, {Wang}, {Wang}, {Wang}, {Wang}, {Wei}, {Wei}, {Wei}, {Wen}, {Wu}, {Wu}, {Wu}, {Wu}, {Wu}, {Xi}, {Xia}, {Xia}, {Xiang}, {Xiao}, {Xiao}, {Xin}, {Xin}, {Xing}, {Xu}, {Xu}, {Xue}, {Yan}, {Yang}, {Yang}, {Yang}, {Yang}, {Yang}, {Yang}, {Yang}, {Yao}, {Yao}, {Ye}, {Yin}, {Yin}, {You}, {You}, {Yu}, {Yuan}, {Zeng}, {Zeng}, {Zeng}, {Zeng}, {Zha}, {Zhai}, {Zhang}, {Zhang}, {Zhang}, {Zhang}, {Zhang}, {Zhang}, {Zhang}, {Zhang}, {Zhang}, {Zhang}, {Zhang}, {Zhang}, {Zhang}, {Zhang}, {Zhang}, {Zhang}, {Zhang}, {Zhang}, {Zhang}, {Zhao}, {Zhao}, {Zhao}, {Zhao}, {Zhao}, {Zheng}, {Zheng}, {Zhou}, {Zhou}, {Zhou}, {Zhou}, {Zhou}, {Zhou}, {Zhu}, {Zhu},
  {Zhu}, {Zhu}, {Zuo}, \& {Collaboration)}}]{aharonian21_lhaasocrab}
{Aharonian}, F., {An}, Q., {Axikegu}, {et~al.} 2021, Chinese Physics C, 45, 085002

\bibitem[{{Aharonian} {et~al.}(1999){Aharonian}, {Akhperjanian}, {Barrio}, {Bernl{\"o}hr}, {Bojahr}, {Calle}, {Contreras}, {Cortina}, {Daum}, {Deckers}, {Denninghoff}, {Fonseca}, {Gonzalez}, {Heinzelmann}, {Hemberger}, {Hermann}, {He{\ss}}, {Heusler}, {Hofmann}, {Hohl}, {Horns}, {Ibarra}, {Kankanyan}, {Kettler}, {K{\"o}hler}, {Konopelko}, {Kornmeyer}, {Kestel}, {Kranich}, {Krawczynski}, {Lampeitl}, {Lindner}, {Lorenz}, {Magnussen}, {Meyer}, {Mirzoyan}, {Moralejo}, {Padilla}, {Panter}, {Petry}, {Plaga}, {Plyasheshnikov}, {Prahl}, {P{\"u}hlhofer}, {Rauterberg}, {Renault}, {Rhode}, {R{\"o}hring}, {Sahakian}, {Samorski}, {Schmele}, {Schr{\"o}der}, {Stamm}, {V{\"o}lk}, {Wiebel-Sooth}, {Wiedner}, {Willmer}, \& {Wittek}}]{aharonian99}
{Aharonian}, F.~A., {Akhperjanian}, A.~G., {Barrio}, J.~A., {et~al.} 1999, \aap, 349, 11

\bibitem[{{Ajello} {et~al.}(2017){Ajello}, {Atwood}, {Baldini}, {Ballet}, {Barbiellini}, {Bastieri}, {Bellazzini}, {Bissaldi}, {Blandford}, {Bloom}, {Bonino}, {Bregeon}, {Britto}, {Bruel}, {Buehler}, {Buson}, {Cameron}, {Caputo}, {Caragiulo}, {Caraveo}, {Cavazzuti}, {Cecchi}, {Charles}, {Chekhtman}, {Cheung}, {Chiaro}, {Ciprini}, {Cohen}, {Costantin}, {Costanza}, {Cuoco}, {Cutini}, {D'Ammando}, {de Palma}, {Desiante}, {Digel}, {Di Lalla}, {Di Mauro}, {Di Venere}, {Dom{\'\i}nguez}, {Drell}, {Dumora}, {Favuzzi}, {Fegan}, {Ferrara}, {Fortin}, {Franckowiak}, {Fukazawa}, {Funk}, {Fusco}, {Gargano}, {Gasparrini}, {Giglietto}, {Giommi}, {Giordano}, {Giroletti}, {Glanzman}, {Green}, {Grenier}, {Grondin}, {Grove}, {Guillemot}, {Guiriec}, {Harding}, {Hays}, {Hewitt}, {Horan}, {J{\'o}hannesson}, {Kensei}, {Kuss}, {La Mura}, {Larsson}, {Latronico}, {Lemoine-Goumard}, {Li}, {Longo}, {Loparco}, {Lott}, {Lubrano}, {Magill}, {Maldera}, {Manfreda}, {Mazziotta}, {McEnery}, {Meyer}, {Michelson}, {Mirabal}, {Mitthumsiri},
  {Mizuno}, {Moiseev}, {Monzani}, {Morselli}, {Moskalenko}, {Negro}, {Nuss}, {Ohsugi}, {Omodei}, {Orienti}, {Orlando}, {Palatiello}, {Paliya}, {Paneque}, {Perkins}, {Persic}, {Pesce-Rollins}, {Piron}, {Porter}, {Principe}, {Rain{\`o}}, {Rando}, {Razzano}, {Razzaque}, {Reimer}, {Reimer}, {Reposeur}, {Saz Parkinson}, {Sgr{\`o}}, {Simone}, {Siskind}, {Spada}, {Spandre}, {Spinelli}, {Stawarz}, {Suson}, {Takahashi}, {Tak}, {Thayer}, {Thayer}, {Thompson}, {Torres}, {Torresi}, {Troja}, {Vianello}, {Wood}, \& {Wood}}]{ajell017_3fhl}
{Ajello}, M., {Atwood}, W.~B., {Baldini}, L., {et~al.} 2017, \apjs, 232, 18

\bibitem[{{Albert} {et~al.}(2020){Albert}, {Alfaro}, {Alvarez}, {Angeles Camacho}, {Arteaga-Vel{\'a}zquez}, {Arunbabu}, {Avila Rojas}, {Ayala Solares}, {Baghmanyan}, {Belmont-Moreno}, {BenZvi}, {Brisbois}, {Caballero-Mora}, {Capistr{\'a}n}, {Carrami{\~n}ana}, {Casanova}, {Cotti}, {Cotzomi}, {Couti{\~n}o de Le{\'o}n}, {De la Fuente}, {de Le{\'o}n}, {Dingus}, {DuVernois}, {D{\'\i}az-V{\'e}lez}, {Ellsworth}, {Engel}, {Espinoza}, {Fleischhack}, {Fraija}, {Galv{\'a}n-G{\'a}mez}, {Garcia}, {Garc{\'\i}a-Gonz{\'a}lez}, {Garfias}, {Gonz{\'a}lez}, {Goodman}, {Harding}, {Hernandez}, {Hona}, {Huang}, {Hueyotl-Zahuantitla}, {H{\"u}ntemeyer}, {Iriarte}, {Joshi}, {Lara}, {Lee}, {Le{\'o}n Vargas}, {Linnemann}, {Longinotti}, {Luis-Raya}, {Lundeen}, {L{\'o}pez-Coto}, {Malone}, {Marinelli}, {Martinez-Castellanos}, {Mart{\'\i}nez-Castro}, {Mart{\'\i}nez-Huerta}, {Matthews}, {Miranda-Romagnoli}, {Morales-Soto}, {Moreno}, {Nayerhoda}, {Nellen}, {Newbold}, {Nisa}, {Noriega-Papaqui}, {Omodei}, {Peisker}, {P{\'e}rez-P{\'e}rez},
  {Rho}, {Rivi{\`e}re}, {Rosa-Gonz{\'a}lez}, {Rosenberg}, {Ruiz-Velasco}, {Salazar}, {Salesa Greus}, {Sandoval}, {Schneider}, {Schoorlemmer}, {Sinnis}, {Smith}, {Springer}, {Surajbali}, {Tabachnick}, {Tanner}, {Tibolla}, {Tollefson}, {Torres}, {Torres-Escobedo}, {Weisgarber}, {Yodh}, {Zepeda}, {Zhou}, \& {HAWC Collaboration}}]{albert20}
{Albert}, A., {Alfaro}, R., {Alvarez}, C., {et~al.} 2020, \prl, 124, 131101

\bibitem[{{Amelino-Camelia} {et~al.}(1998){Amelino-Camelia}, {Ellis}, {Mavromatos}, \& {Sarkar}}]{amelino98}
{Amelino-Camelia}, G., {Ellis}, J., {Mavromatos}, N.~E. a nd~{Nanopoulos}, D.~V., \& {Sarkar}, S. 1998, \nat, 393, 763

\bibitem[{{Amelino-Camelia} \& {Piran}(2001)}]{amelino01}
{Amelino-Camelia}, G., \& {Piran}, T. 2001, \prd, 64, 036005

\bibitem[{{Andrews} {et~al.}(2018){Andrews}, {Driver}, {Davies}, {Lagos}, \& {Robotham}}]{andrews18}
{Andrews}, S.~K., {Driver}, S.~P., {Davies}, L.~J.~M., {Lagos}, C. d.~P., \& {Robotham}, A.~S.~G. 2018, \mnras, 474, 898

\bibitem[{{Baktash} {et~al.}(2022){Baktash}, {Horns}, \& {Meyer}}]{baktash22}
{Baktash}, A., {Horns}, D., \& {Meyer}, M. 2022, arXiv e-prints, arXiv:2210.07172

\bibitem[{{Balaji}(2023)}]{balaji23}
{Balaji}, S. 2023, Physics Letters B, 845, 138157

\bibitem[{{Biteau} \& {Williams}(2015)}]{biteau15}
{Biteau}, J., \& {Williams}, D.~A. 2015, \apj, 812, 60

\bibitem[{{Boettcher} \& {Dermer}(1995)}]{boettcher95}
{Boettcher}, M., \& {Dermer}, C.~D. 1995, \aap, 302, 37

\bibitem[{{Brown} {et~al.}(1973){Brown}, {Hunt}, {Mikaelian}, \& {Muzinich}}]{brown73_PRD}
{Brown}, R.~W., {Hunt}, W.~F., {Mikaelian}, K.~O., \& {Muzinich}, I.~J. 1973, \prd, 8, 3083

\bibitem[{{Cao} {et~al.}(2019){Cao}, {della Volpe}, {Liu}, {Editors}, {:}, {Bi}, {Chen}, {D'Ettorre Piazzoli}, {Feng}, {Jia}, {Li}, {Ma}, {Wang}, {Zhang}, {Referees}, {:}, {Qie}, {Hu}, {Referees}, {:}, {S{\'a}iz}, {Yang}, {Contributors}, {:}, {Addazi}, {Belotsky}, {Beylin}, {Bi}, {Che}, {Chen}, {Cheng}, {Chiavassa}, {Cirelli}, {Di Sciascio}, {Esmaili}, {Fang}, {Fornengo}, {Gou}, {Guo}, {Gan}, {Gong}, {Gu}, {He}, {He}, {Hou}, {Huang}, {Huang}, {Kachekriess}, {Khlopov}, {Korchagin}, {Korochkin}, {Kuksa}, {Ksenofontov}, {Liu}, {Liu}, {Liu}, {Marciano}, {Martineau-Huynh}, {Martraire}, {Ma}, {Neronov}, {Panci}, {Pasechnick}, {Ruffolo}, {Sakharov}, {Sala}, {Semikoz}, {Shchegolev}, {Serpico}, {Sheng}, {Stenkin}, {Tam}, {Vernetto}, {Vallania}, {Volchanskiy}, {Wang}, {Wang}, {Wang}, {Wu}, {Wu}, {Wu}, {Xiao}, {Yang}, {Yan}, {Yao}, {Yin}, {Yuan}, {Zhang}, {Zeng}, {Zhang}, {Zhang}, {Zhou}, {Zhu}, \& {Zuo}}]{cao19}
{Cao}, Z., {della Volpe}, D., {Liu}, S., {et~al.} 2019, arXiv e-prints, arXiv:1905.02773

\bibitem[{{Cao} {et~al.}(2021{\natexlab{a}}){Cao}, {Aharonian}, {An}, {Axikegu}, {Bai}, {Bai}, {Bao}, {Bastieri}, {Bi}, {Bi}, {Cai}, {Cai}, {Cao}, {Chang}, {Chang}, {Chen}, {Chen}, {Chen}, {Chen}, {Chen}, {Chen}, {Chen}, {Chen}, {Chen}, {Chen}, {Chen}, {Chen}, {Chen}, {Chen}, {Cheng}, {Cheng}, {Cui}, {Cui}, {Cui}, {D'Ettorre Piazzoli}, {Dai}, {Dai}, {Dai}, {Danzengluobu}, {Della Volpe}, {Dong}, {Duan}, {Fan}, {Fan}, {Fan}, {Fang}, {Fang}, {Feng}, {Feng}, {Feng}, {Feng}, {Gao}, {Gao}, {Gao}, {Gao}, {Gao}, {Ge}, {Geng}, {Gong}, {Gou}, {Gu}, {Guo}, {Guo}, {Guo}, {Guo}, {Guo}, {Han}, {He}, {He}, {He}, {He}, {He}, {He}, {Heller}, {Hor}, {Hou}, {Hou}, {Hu}, {Hu}, {Hu}, {Hu}, {Huang}, {Huang}, {Huang}, {Huang}, {Huang}, {Huang}, {Ji}, {Ji}, {Jia}, {Jiang}, {Jiang}, {Jin}, {Ke}, {Kuleshov}, {Levochkin}, {Li}, {Li}, {Li}, {Li}, {Li}, {Li}, {Li}, {Li}, {Li}, {Li}, {Li}, {Li}, {Li}, {Li}, {Li}, {Li}, {Li}, {Li}, {Liang}, {Liang}, {Lin}, {Liu}, {Liu}, {Liu}, {Liu}, {Liu}, {Liu}, {Liu}, {Liu}, {Liu}, {Liu}, {Liu}, {Liu},
  {Liu}, {Liu}, {Liu}, {Liu}, {Long}, {Lu}, {Lv}, {Ma}, {Ma}, {Ma}, {Mao}, {Masood}, {Min}, {Mitthumsiri}, {Montaruli}, {Nan}, {Pang}, {Pattarakijwanich}, {Pei}, {Qi}, {Qi}, {Qiao}, {Qin}, {Ruffolo}, {Rulev}, {Saiz}, {Shao}, {Shchegolev}, {Sheng}, {Shi}, {Song}, {Stenkin}, {Stepanov}, {Su}, {Sun}, {Sun}, {Sun}, {Tam}, {Tang}, {Tian}, {Wang}, {Wang}, {Wang}, {Wang}, {Wang}, {Wang}, {Wang}, {Wang}, {Wang}, {Wang}, {Wang}, {Wang}, {Wang}, {Wang}, {Wang}, {Wang}, {Wang}, {Wang}, {Wang}, {Wang}, {Wang}, {Wang}, {Wei}, {Wei}, {Wei}, {Wen}, {Wu}, {Wu}, {Wu}, {Wu}, {Wu}, {Xi}, {Xia}, {Xia}, {Xiang}, {Xiao}, {Xiao}, {Xiao}, {Xin}, {Xin}, {Xing}, {Xu}, {Xu}, {Xue}, {Yan}, {Yan}, {Yang}, {Yang}, {Yang}, {Yang}, {Yang}, {Yang}, {Yang}, {Yao}, {Yao}, {Ye}, {Yin}, {Yin}, {You}, {You}, {Yu}, {Yuan}, {Zeng}, {Zeng}, {Zeng}, {Zeng}, {Zha}, {Zhai}, {Zhang}, {Zhang}, {Zhang}, {Zhang}, {Zhang}, {Zhang}, {Zhang}, {Zhang}, {Zhang}, {Zhang}, {Zhang}, {Zhang}, {Zhang}, {Zhang}, {Zhang}, {Zhang}, {Zhang}, {Zhang}, {Zhang}, {Zhao},
  {Zhao}, {Zhao}, {Zhao}, {Zhao}, {Zheng}, {Zheng}, {Zhou}, {Zhou}, {Zhou}, {Zhou}, {Zhou}, {Zhou}, {Zhu}, {Zhu}, {Zhu}, {Zhu}, \& {Zuo}}]{cao21_lhaasocrab}
{Cao}, Z., {Aharonian}, F., {An}, Q., {et~al.} 2021{\natexlab{a}}, Science, 373, 425

\bibitem[{{Cao} {et~al.}(2021{\natexlab{b}}){Cao}, {Aharonian}, {An}, {Axikegu}, {Bai}, {Bao}, {Bastieri}, {Bi}, {Bi}, {Cai}, {Cai}, {Cao}, {Chang}, {Chang}, {Chang}, {Chen}, {Chen}, {Chen}, {Chen}, {Chen}, {Chen}, {Chen}, {Chen}, {Chen}, {Chen}, {Chen}, {Chen}, {Chen}, {Cheng}, {Cheng}, {Cui}, {Cui}, {Cui}, {Dai}, {Dai}, {Dai}, {Danzengluobu}, {della Volpe}, {D'Ettorre Piazzoli}, {Dong}, {Fan}, {Fan}, {Fan}, {Fang}, {Fang}, {Feng}, {Feng}, {Feng}, {Feng}, {Gao}, {Gao}, {Gao}, {Gao}, {Ge}, {Geng}, {Gong}, {Gou}, {Gu}, {Guo}, {Guo}, {Guo}, {Guo}, {Han}, {He}, {He}, {He}, {He}, {He}, {He}, {Heller}, {Hor}, {Hou}, {Hou}, {Hu}, {Hu}, {Hu}, {Hu}, {Huang}, {Huang}, {Huang}, {Huang}, {Huang}, {Ji}, {Ji}, {Jia}, {Jiang}, {Jiang}, {Jin}, {Kuleshov}, {Levochkin}, {Li}, {Li}, {Li}, {Li}, {Li}, {Li}, {Li}, {Li}, {Li}, {Li}, {Li}, {Li}, {Li}, {Li}, {Li}, {Li}, {Li}, {Liang}, {Liang}, {Lin}, {Liu}, {Liu}, {Liu}, {Liu}, {Liu}, {Liu}, {Liu}, {Liu}, {Liu}, {Liu}, {Liu}, {Liu}, {Liu}, {Liu}, {Liu}, {Long}, {Lu}, {Lv}, {Ma},
  {Ma}, {Ma}, {Mao}, {Masood}, {Mitthumsiri}, {Montaruli}, {Nan}, {Pang}, {Pattarakijwanich}, {Pei}, {Qi}, {Ruffolo}, {Rulev}, {S{\'a}iz}, {Shao}, {Shchegolev}, {Sheng}, {Shi}, {Song}, {Stenkin}, {Stepanov}, {Sun}, {Sun}, {Sun}, {Tam}, {Tang}, {Tian}, {Wang}, {Wang}, {Wang}, {Wang}, {Wang}, {Wang}, {Wang}, {Wang}, {Wang}, {Wang}, {Wang}, {Wang}, {Wang}, {Wang}, {Wang}, {Wang}, {Wang}, {Wang}, {Wang}, {Wang}, {Wang}, {Wei}, {Wei}, {Wei}, {Wen}, {Wu}, {Wu}, {Wu}, {Wu}, {Wu}, {Xi}, {Xia}, {Xia}, {Xiang}, {Xiao}, {Xiao}, {Xin}, {Xin}, {Xing}, {Xu}, {Xu}, {Xue}, {Yan}, {Yang}, {Yang}, {Yang}, {Yang}, {Yang}, {Yang}, {Yang}, {Yao}, {Yao}, {Ye}, {Yin}, {Yin}, {You}, {You}, {Yu}, {Yuan}, {Zeng}, {Zeng}, {Zeng}, {Zeng}, {Zha}, {Zhai}, {Zhang}, {Zhang}, {Zhang}, {Zhang}, {Zhang}, {Zhang}, {Zhang}, {Zhang}, {Zhang}, {Zhang}, {Zhang}, {Zhang}, {Zhang}, {Zhang}, {Zhang}, {Zhang}, {Zhang}, {Zhang}, {Zhang}, {Zhao}, {Zhao}, {Zhao}, {Zhao}, {Zhao}, {Zheng}, {Zheng}, {Zhou}, {Zhou}, {Zhou}, {Zhou}, {Zhou}, {Zhou}, {Zhu},
  {Zhu}, {Zhu}, {Zhu}, \& {Zuo}}]{cao21}
{Cao}, Z., {Aharonian}, F.~A., {An}, Q., {et~al.} 2021{\natexlab{b}}, \nat, 594, 33

\bibitem[{{Cao} {et~al.}(2023){Cao}, {Aharonian}, {An}, {Axikegu}, {Bai}, {Bao}, {Bastieri}, {Bi}, {Bi}, {Cai}, {Cao}, {Cao}, {Cao}, {Chang}, {Chang}, {Chen}, {Chen}, {Chen}, {Chen}, {Chen}, {Chen}, {Chen}, {Chen}, {Chen}, {Chen}, {Chen}, {Chen}, {Cheng}, {Cheng}, {Cui}, {Cui}, {Cui}, {Cui}, {Dai}, {Dai}, {Dai}, {Danzengluobu}, {della Volpe}, {Dong}, {Duan}, {Fan}, {Fan}, {Fang}, {Fang}, {Feng}, {Feng}, {Feng}, {Feng}, {Feng}, {Gabici}, {Gao}, {Gao}, {Gao}, {Gao}, {Gao}, {Gao}, {Ge}, {Geng}, {Giacinti}, {Gong}, {Gou}, {Gu}, {Guo}, {Guo}, {Guo}, {Guo}, {Han}, {He}, {He}, {He}, {He}, {He}, {Heller}, {Hor}, {Hou}, {Hou}, {Hou}, {Hu}, {Hu}, {Hu}, {Huang}, {Huang}, {Huang}, {Huang}, {Huang}, {Huang}, {Huang}, {Ji}, {Jia}, {Jia}, {Jiang}, {Jiang}, {Jiang}, {Jin}, {Kang}, {Ke}, {Kuleshov}, {Kurinov}, {Li}, {Li}, {Li}, {Li}, {Li}, {Li}, {Li}, {Li}, {Li}, {Li}, {Li}, {Li}, {Li}, {Li}, {Li}, {Li}, {Li}, {Li}, {Li}, {Liang}, {Liang}, {Lin}, {Liu}, {Liu}, {Liu}, {Liu}, {Liu}, {Liu}, {Liu}, {Liu}, {Liu}, {Liu}, {Liu},
  {Liu}, {Liu}, {Liu}, {Lu}, {Luo}, {Lv}, {Ma}, {Ma}, {Ma}, {Mao}, {Min}, {Mitthumsiri}, {Mu}, {Nan}, {Neronov}, {Ou}, {Pang}, {Pattarakijwanich}, {Pei}, {Qi}, {Qi}, {Qiao}, {Qin}, {Ruffolo}, {S{\'a}iz}, {Semikoz}, {Shao}, {Shao}, {Shchegolev}, {Sheng}, {Shu}, {Song}, {Stenkin}, {Stepanov}, {Su}, {Sun}, {Sun}, {Sun}, {Tam}, {Tang}, {Tang}, {Tian}, {Wang}, {Wang}, {Wang}, {Wang}, {Wang}, {Wang}, {Wang}, {Wang}, {Wang}, {Wang}, {Wang}, {Wang}, {Wang}, {Wang}, {Wang}, {Wang}, {Wang}, {Wang}, {Wang}, {Wang}, {Wang}, {Wei}, {Wei}, {Wei}, {Wen}, {Wu}, {Wu}, {Wu}, {Wu}, {Wu}, {Xi}, {Xia}, {Xia}, {Xiang}, {Xiao}, {Xiao}, {Xin}, {Xin}, {Xing}, {Xiong}, {Xu}, {Xu}, {Xu}, {Xu}, {Xue}, {Yan}, {Yan}, {Yan}, {Yang}, {Yang}, {Yang}, {Yang}, {Yang}, {Yang}, {Yang}, {Yang}, {Yang}, {Yao}, {Yao}, {Ye}, {Yin}, {Yin}, {You}, {You}, {Yu}, {Yuan}, {Yue}, {Zeng}, {Zeng}, {Zeng}, {Zha}, {Zhang}, {Zhang}, {Zhang}, {Zhang}, {Zhang}, {Zhang}, {Zhang}, {Zhang}, {Zhang}, {Zhang}, {Zhang}, {Zhang}, {Zhang}, {Zhang}, {Zhang}, {Zhang},
  {Zhang}, {Zhang}, {Zhao}, {Zhao}, {Zhao}, {Zhao}, {Zhao}, {Zheng}, {Zhou}, {Zhou}, {Zhou}, {Zhou}, {Zhou}, {Zhou}, {Zhou}, {Zhu}, {Zhu}, {Zhu}, {Zhu}, \& {Zuo.}}]{cao23}
{Cao}, Z., {Aharonian}, F., {An}, Q., {et~al.} 2023, arXiv e-prints, arXiv:2305.17030

\bibitem[{{Christiansen} {et~al.}(2006){Christiansen}, {Ng}, \& {van Dam}}]{christiansen06}
{Christiansen}, W.~A., {Ng}, Y.~J., \& {van Dam}, H. 2006, \prl, 96, 051301

\bibitem[{{Dermer} {et~al.}(2009){Dermer}, {Finke}, {Krug}, \& {B{\"o}ttcher}}]{dermer09}
{Dermer}, C.~D., {Finke}, J.~D., {Krug}, H., \& {B{\"o}ttcher}, M. 2009, \apj, 692, 32

\bibitem[{{Dermer} \& {Menon}(2009)}]{dermer09_book}
{Dermer}, C.~D., \& {Menon}, G. 2009, {High Energy Radiation from Black Holes: Gamma Rays, Cosmic Rays, and Neutrinos}

\bibitem[{{Desai}(2023)}]{desai23}
{Desai}, S. 2023, arXiv e-prints, arXiv:2303.10643

\bibitem[{{DeYoung} \& {HAWC Collaboration}(2012)}]{deyoung12}
{DeYoung}, T., \& {HAWC Collaboration}. 2012, Nuclear Instruments and Methods in Physics Research A, 692, 72

\bibitem[{{Dom{\'{\i}}nguez} {et~al.}(2011)}]{dominguez11}
{Dom{\'{\i}}nguez}, A., {et~al.} 2011, \mnras, 410, 2556

\bibitem[{{Dzhappuev} {et~al.}(2022){Dzhappuev}, {Afashokov}, {Dzaparova}, {Dzhatdoev}, {Gorbacheva}, {Karpikov}, {Khadzhi ev}, {Klimenko}, {Kudzhaev}, {Kurenya}, {L idvansky}, {Mikhailova}, {Petkov}, {Podlesnyi}, {Pozdnukhov}, {Romanenko}, {Rubtsov}, {Troitsk y}, {Unatlokov}, {Vaiman}, {Yanin}, \& {Zhura vleva}}]{dzhappuev22}
{Dzhappuev}, D.~D., {Afashokov}, Y.~Z., {Dzaparova}, I.~M., {et~al.} 2022, The Astronomer's Telegram, 15669, 1

\bibitem[{{Ellis} {et~al.}(2019){Ellis}, {Konoplich}, {Mavromatos}, {Nguyen}, {Sakharov}, \& {Sarkisyan-Grinbaum}}]{ellis19}
{Ellis}, J., {Konoplich}, R., {Mavromatos}, N.~E., {et~al.} 2019, \prd, 99, 083009

\bibitem[{{Ellis} {et~al.}(2008){Ellis}, {Mavromatos}, \& {Nanopoulos}}]{ellis08}
{Ellis}, J., {Mavromatos}, N.~E., \& {Nanopoulos}, D.~V. 2008, Physics Letters B, 665, 412

\bibitem[{{Fairbairn} {et~al.}(2014){Fairbairn}, {Nilsson}, {Ellis}, {Hinton}, \& {White}}]{fairbairn14}
{Fairbairn}, M., {Nilsson}, A., {Ellis}, J., {Hinton}, J., \& {White}, R. 2014, \jcap, 2014, 005

\bibitem[{{Fazio} \& {Stecker}(1970)}]{fazio70}
{Fazio}, G.~G., \& {Stecker}, F.~W. 1970, \nat, 226, 135

\bibitem[{{Finke}(2016)}]{finke16}
{Finke}, J.~D. 2016, \apj, 830, 94

\bibitem[{{Finke} {et~al.}(2022){Finke}, {Ajello}, {Dom{\'\i}nguez}, {Desai}, {Hartmann}, {Paliya}, \& {Saldana-Lopez}}]{finke22}
{Finke}, J.~D., {Ajello}, M., {Dom{\'\i}nguez}, A., {et~al.} 2022, \apj, 941, 33

\bibitem[{{Finke} \& {Razzaque}(2023)}]{finke23}
{Finke}, J.~D., \& {Razzaque}, S. 2023, \apjl, 942, L21

\bibitem[{{Finke} {et~al.}(2010){Finke}, {Razzaque}, \& {Dermer}}]{finke10_EBL}
{Finke}, J.~D., {Razzaque}, S., \& {Dermer}, C.~D. 2010, \apj, 712, 238

\bibitem[{{Franceschini} \& {Rodighiero}(2017)}]{franceschini17}
{Franceschini}, A., \& {Rodighiero}, G. 2017, \aap, 603, A34

\bibitem[{{Franceschini} {et~al.}(2008){Franceschini}, {Rodighiero}, \& {Vaccari}}]{franceschini08}
{Franceschini}, A., {Rodighiero}, G., \& {Vaccari}, M. 2008, \aap, 487, 837

\bibitem[{{Gould} \& {Schr{\'e}der}(1967{\natexlab{a}})}]{gould67_EBL}
{Gould}, R.~J., \& {Schr{\'e}der}, G.~P. 1967{\natexlab{a}}, Physical Review, 155, 1408

\bibitem[{{Gould} \& {Schr{\'e}der}(1967{\natexlab{b}})}]{gould67}
---. 1967{\natexlab{b}}, Physical Review, 155, 1404

\bibitem[{{Guedes Lang} {et~al.}(2018){Guedes Lang}, {Mart{\'\i}nez-Huerta}, \& {de Souza}}]{lang18}
{Guedes Lang}, R., {Mart{\'\i}nez-Huerta}, H., \& {de Souza}, V. 2018, \apj, 853, 23

\bibitem[{{Hauser} \& {Dwek}(2001)}]{hauser01}
{Hauser}, M.~G., \& {Dwek}, E. 2001, \araa, 39, 249

\bibitem[{{Helgason} \& {Kashlinsky}(2012)}]{helgason12}
{Helgason}, K., \& {Kashlinsky}, A. 2012, \apjl, 758, L13

\bibitem[{{Inoue} \& {Tanaka}(2016)}]{inoue16}
{Inoue}, Y., \& {Tanaka}, Y.~T. 2016, \apj, 818, 187

\bibitem[{{Jacob} \& {Piran}(2008)}]{jacob08}
{Jacob}, U., \& {Piran}, T. 2008, \prd, 78, 124010

\bibitem[{{Jacobson} {et~al.}(2006){Jacobson}, {Liberati}, \& {Mattingly}}]{jacobson06}
{Jacobson}, T., {Liberati}, S., \& {Mattingly}, D. 2006, Annals of Physics, 321, 150

\bibitem[{{Khaire} \& {Srianand}(2015)}]{khaire15}
{Khaire}, V., \& {Srianand}, R. 2015, \apj, 805, 33

\bibitem[{{Khaire} \& {Srianand}(2019)}]{khaire19}
---. 2019, \mnras, 484, 4174

\bibitem[{{Kifune}(1999)}]{kifune99}
{Kifune}, T. 1999, \apjl, 518, L21

\bibitem[{{Kneiske} \& {Dole}(2010)}]{kneiske10}
{Kneiske}, T.~M., \& {Dole}, H. 2010, \aap, 515, A19

\bibitem[{{Li} \& {Ma}(2023)}]{li23}
{Li}, H., \& {Ma}, B.-Q. 2023, Astroparticle Physics, 148, 102831

\bibitem[{{Linden} {et~al.}(2022){Linden}, {Beacom}, {Peter}, {Buckman}, {Zhou}, \& {Zhu}}]{linden22}
{Linden}, T., {Beacom}, J.~F., {Peter}, A. H.~G., {et~al.} 2022, \prd, 105, 063013

\bibitem[{{Loeb}(2022)}]{loeb22}
{Loeb}, A. 2022, Research Notes of the American Astronomical Society, 6, 148

\bibitem[{{Mart{\'\i}nez-Huerta} {et~al.}(2020){Mart{\'\i}nez-Huerta}, {Lang}, \& { de Souza}}]{martinez20}
{Mart{\'\i}nez-Huerta}, H., {Lang}, R.~G., \& { de Souza}, V. 2020, Symmetry, 12, 1232

\bibitem[{{Mattingly}(2005)}]{mattingly05}
{Mattingly}, D. 2005, Living Reviews in Relativity, 8, 5

\bibitem[{{Moskalenko} {et~al.}(2006){Moskalenko}, {Porter}, \& {Digel}}]{mosk06}
{Moskalenko}, I.~V., {Porter}, T.~A., \& {Digel}, S.~W. 2006, \apjl, 652, L65

\bibitem[{{Nikishov}(1962)}]{nikishov62}
{Nikishov}, A.~I. 1962, JETP, 393, 14

\bibitem[{{Orlando} \& {Strong}(2007)}]{orlando07}
{Orlando}, E., \& {Strong}, A.~W. 2007, \apss, 309, 359

\bibitem[{{Orlando} \& {Strong}(2008)}]{orlando08}
---. 2008, \aap, 480, 847

\bibitem[{{Protheroe} \& {Meyer}(2000)}]{protheroe00}
{Protheroe}, R.~J., \& {Meyer}, H. 2000, Physics Letters B, 493, 1

\bibitem[{{Qu} {et~al.}(2019){Qu}, {Zeng}, \& {Yan}}]{qu19}
{Qu}, Y., {Zeng}, H., \& {Yan}, D. 2019, \mnras, 490, 758

\bibitem[{{Qu} \& {Zeng}(2022)}]{qu22}
{Qu}, Y.-k., \& {Zeng}, H.-d. 2022, \caa, 46, 42

\bibitem[{{Razzaque} {et~al.}(2009){Razzaque}, {Dermer}, \& {Finke}}]{razzaque09}
{Razzaque}, S., {Dermer}, C.~D., \& {Finke}, J.~D. 2009, \apj, 697, 483

\bibitem[{{Saldana-Lopez} {et~al.}(2021){Saldana-Lopez}, {Dom{\'\i}nguez}, {P{\'e}rez-Gonz{\'a}lez}, {Finke}, {Ajello}, {Primack}, {Paliya}, \& {Desai}}]{saldana21}
{Saldana-Lopez}, A., {Dom{\'\i}nguez}, A., {P{\'e}rez-Gonz{\'a}lez}, P.~G., {et~al.} 2021, \mnras, 507, 5144

\bibitem[{{Sarkar}(2002)}]{sarkar02}
{Sarkar}, S. 2002, Modern Physics Letters A, 17, 1025

\bibitem[{{Scully} {et~al.}(2014){Scully}, {Malkan}, \& {Stecker}}]{scully14}
{Scully}, S.~T., {Malkan}, M.~A., \& {Stecker}, F.~W. 2014, \apj, 784, 138

\bibitem[{{Stecker} \& {Glashow}(2001)}]{stecker01}
{Stecker}, F.~W., \& {Glashow}, S.~L. 2001, Astroparticle Physics, 16, 97

\bibitem[{{Stecker} {et~al.}(2012){Stecker}, {Malkan}, \& {Scully}}]{stecker12}
{Stecker}, F.~W., {Malkan}, M.~A., \& {Scully}, S.~T. 2012, \apj, 761, 128

\bibitem[{{Stecker} {et~al.}(2016){Stecker}, {Scully}, \& {Malkan}}]{stecker16}
{Stecker}, F.~W., {Scully}, S.~T., \& {Malkan}, M.~A. 2016, \apj, 827, 6

\bibitem[{{Tanabashi} {et~al.}(2018){Tanabashi}, {Hagiwara}, {Hikasa}, {Nakamura}, {Sumino}, {Takahashi}, {Tanaka}, {Agashe}, {Aielli}, {Amsler}, {Antonelli}, {Asner}, {Baer}, {Banerjee}, {Barnett}, {Basaglia}, {Bauer}, {Beatty}, {Belousov}, {Beringer}, {Bethke}, {Bettini}, {Bichsel}, {Biebel}, {Black}, {Blucher}, {Buchmuller}, {Burkert}, {Bychkov}, {Cahn}, {Carena}, {Ceccucci}, {Cerri}, {Chakraborty}, {Chen}, {Chivukula}, {Cowan}, {Dahl}, {D'Ambrosio}, {Damour}, {de Florian}, {de Gouv{\^e}a}, {DeGrand}, {de Jong}, {Dissertori}, {Dobrescu}, {D'Onofrio}, {Doser}, {Drees}, {Dreiner}, {Dwyer}, {Eerola}, {Eidelman}, {Ellis}, {Erler}, {Ezhela}, {Fetscher}, {Fields}, {Firestone}, {Foster}, {Freitas}, {Gallagher}, {Garren}, {Gerber}, {Gerbier}, {Gershon}, {Gershtein}, {Gherghetta}, {Godizov}, {Goodman}, {Grab}, {Gritsan}, {Grojean}, {Groom}, {Gr{\"u}newald}, {Gurtu}, {Gutsche}, {Haber}, {Hanhart}, {Hashimoto}, {Hayato}, {Hayes}, {Hebecker}, {Heinemeyer}, {Heltsley}, {Hern{\'a}ndez-Rey}, {Hisano}, {H{\"o}cker},
  {Holder}, {Holtkamp}, {Hyodo}, {Irwin}, {Johnson}, {Kado}, {Karliner}, {Katz}, {Klein}, {Klempt}, {Kowalewski}, {Krauss}, {Kreps}, {Krusche}, {Kuyanov}, {Kwon}, {Lahav}, {Laiho}, {Lesgourgues}, {Liddle}, {Ligeti}, {Lin}, {Lippmann}, {Liss}, {Littenberg}, {Lugovsky}, {Lugovsky}, {Lusiani}, {Makida}, {Maltoni}, {Mannel}, {Manohar}, {Marciano}, {Martin}, {Masoni}, {Matthews}, {Mei{\ss}ner}, {Milstead}, {Mitchell}, {M{\"o}nig}, {Molaro}, {Moortgat}, {Moskovic}, {Murayama}, {Narain}, {Nason}, {Navas}, {Neubert}, {Nevski}, {Nir}, {Olive}, {Pagan Griso}, {Parsons}, {Patrignani}, {Peacock}, {Pennington}, {Petcov}, {Petrov}, {Pianori}, {Piepke}, {Pomarol}, {Quadt}, {Rademacker}, {Raffelt}, {Ratcliff}, {Richardson}, {Ringwald}, {Roesler}, {Rolli}, {Romaniouk}, {Rosenberg}, {Rosner}, {Rybka}, {Ryutin}, {Sachrajda}, {Sakai}, {Salam}, {Sarkar}, {Sauli}, {Schneider}, {Scholberg}, {Schwartz}, {Scott}, {Sharma}, {Sharpe}, {Shutt}, {Silari}, {Sj{\"o}strand}, {Skands}, {Skwarnicki}, {Smith}, {Smoot}, {Spanier}, {Spieler},
  {Spiering}, {Stahl}, {Stone}, {Sumiyoshi}, {Syphers}, {Terashi}, {Terning}, {Thoma}, {Thorne}, {Tiator}, {Titov}, {Tkachenko}, {T{\"o}rnqvist}, {Tovey}, {Valencia}, {Van de Water}, {Varelas}, {Venanzoni}, {Verde}, {Vincter}, {Vogel}, {Vogt}, {Wakely}, {Walkowiak}, {Walter}, {Wands}, {Ward}, {Wascko}, {Weiglein}, {Weinberg}, {Weinberg}, {White}, {Wiencke}, {Willocq}, {Wohl}, {Womersley}, {Woody}, {Workman}, {Yao}, {Zeller}, {Zenin}, {Zhu}, {Zhu}, {Zimmermann}, {Zyla}, {Anderson}, {Fuller}, {Lugovsky}, {Schaffner}, \& {Particle Data Group}}]{tanabashi18}
{Tanabashi}, M., {Hagiwara}, K., {Hikasa}, K., {et~al.} 2018, \prd, 98, 030001

\bibitem[{{Tavecchio} \& {Bonnoli}(2016)}]{tavecchio16}
{Tavecchio}, F., \& {Bonnoli}, G. 2016, \aap, 585, A25

\bibitem[{{Vasileiou} {et~al.}(2013){Vasileiou}, {Jacholkowska}, {Piron}, {Bolmont}, {Couturier}, {Granot}, {Stecker}, {Cohen-Tanug i}, \& {Longo}}]{vasil13}
{Vasileiou}, V., {Jacholkowska}, A., {Piron}, F., {et~al.} 2013, \prd, 87, 122001

\bibitem[{{Wakely} \& {Horan}(2008)}]{wakely08}
{Wakely}, S.~P., \& {Horan}, D. 2008, in International Cosmic Ray Conference, Vol.~3, International Cosmic Ray Conference, 1341--1344

\end{thebibliography}
\bibliographystyle{apj}

\end{document}